\documentclass[twocolumn,aps,prx,floatfix,superscriptaddress]{revtex4-2}

\usepackage{amsmath,amssymb,amsfonts,graphicx,braket,braket,xcolor,soul}
\usepackage{hyperref}
\usepackage[export]{adjustbox}
\usepackage[normalem]{ulem}
\hypersetup{colorlinks = true}
\usepackage{algorithm}
\usepackage[noend]{algpseudocode}
\usepackage{array}
\algrenewcommand\algorithmicrequire{\textbf{Input:}}
\algrenewcommand\algorithmicensure{\textbf{Output:}}

\definecolor{myblue}{RGB}{222,235,247}
\sethlcolor{myblue}

\begin{document}
	
\title{
Simulating continuous-space systems with quantum-classical wave functions
}

\author{Friederike Metz}
\email{friederike.metz@epfl.ch}
\affiliation{Institute of Physics, École Polytechnique Fédérale de Lausanne (EPFL), CH-1015 Lausanne, Switzerland}
\affiliation{Center for Quantum Science and Engineering, École Polytechnique Fédérale de Lausanne (EPFL), CH-1015 Lausanne, Switzerland}

\author{Gabriel Pescia}
\affiliation{Institute of Physics, École Polytechnique Fédérale de Lausanne (EPFL), CH-1015 Lausanne, Switzerland}
\affiliation{Center for Quantum Science and Engineering, École Polytechnique Fédérale de Lausanne (EPFL), CH-1015 Lausanne, Switzerland}

\author{Giuseppe Carleo}
\affiliation{Institute of Physics, École Polytechnique Fédérale de Lausanne (EPFL), CH-1015 Lausanne, Switzerland}
\affiliation{Center for Quantum Science and Engineering, École Polytechnique Fédérale de Lausanne (EPFL), CH-1015 Lausanne, Switzerland}

\date{\today}

\begin{abstract}
Most non-relativistic interacting quantum many-body systems, such as atomic and molecular ensembles or materials, are naturally described in terms of continuous-space Hamiltonians. The simulation of their ground-state properties on digital quantum computers is challenging because current algorithms require discretization, which usually amounts to choosing a finite basis set, inevitably introducing errors. In this work, we propose an alternative, discretization-free approach that combines classical and quantum resources in a global variational ansatz, optimized using the framework of variational Monte Carlo. We introduce both purely quantum as well as hybrid quantum-classical ansatze and benchmark them on three paradigmatic continuous-space systems that are either very challenging or beyond the reach of current quantum approaches: the one-dimensional quantum rotor model, a system of $^3$He particles in one and two dimensions, and the two-dimensional homogeneous electron gas. We embed relevant constraints such as the antisymmetry of fermionic wave functions directly into the ansatz. Many-body correlations are introduced via backflow transformations represented by parameterized quantum circuits. We demonstrate that the accuracy of the simulation can be systematically improved by increasing the number of circuit parameters and study the effects of shot noise. Furthermore, we show that the hybrid ansatz improves the ground-state energies obtained using the purely classical wave function.
\end{abstract}

\maketitle

\section{\label{sec:intro}Introduction}
The ability to efficiently and accurately simulate quantum many-body systems impacts a variety of scientific areas such as condensed matter physics, materials science, and quantum chemistry~\cite{Georgescu14,Cao19,McArdle20,Bauer20,Yuri23,Cao18}. It has led to the development of several specifically tailored computational methods~\cite{Foulkes01,Eberhard90,schollwock11,Georges96}. 
The majority of these methods leverage classical computing resources to predict physical and chemical properties from the system's Hamiltonian. However, the exponentially growing Hilbert-space dimension of quantum many-body systems inevitably hinders an exact classical treatment. To mitigate this \textit{curse of dimensionality}, it is advantageous to utilize quantum resources, such as digital quantum computers, which have significant potential due to the availability of quantum simulation algorithms that scale polynomially with the size of the system~\cite{Georgescu14,Lloyd1996}.

Since the natural description of most systems of interest include continuous degrees of freedom~\cite{Cao19,McArdle20} (e.g. particle positions), the usage of (discrete) quantum computing resources poses a challenge, that is subject to severe bottlenecks. 
Typically, one reformulates the continuous problem in \textit{second quantization} by introducing an (infinite) set of basis functions, such as plane waves or Gaussian functions, that is truncated to a finite size. The truncation inevitably introduces errors, which can only be quantified by systematically studying the effects of extending the set of basis functions and, therefore, increasing the number of qubits and measurements on the quantum device.

Another possible route are \textit{first-quantized} approaches that are based on, for example, a discrete real-space grid and the split-operator method \cite{Zalka69,Kassal08,Kassal11,Hans23}, a configuration interaction representation \cite{Babbush17,toloui2013}, the qubitization technique \cite{Low2019,Berry18}, or an interaction picture algorithm \cite{Babbush19,Su21}. 
Although still intrinsically discretized, these provide a more favorable asymptotic scaling in both the number of particles and the number of orbitals than second-quantized techniques, but require fault tolerance, which is still beyond the reach of near-term hardware \cite{Babbush19,Su21,Beverland22}.

To harness the currently available quantum devices for large-scale quantum simulations, a possible remedy is the combination of quantum and classical techniques \cite{Bravyi16,Peng20,Mitarai21,Piveteau24,Gian,Eddins22,huembeli22,Schoulepnikoff24,Bauer16,rubin2016,Rossmannek21,Rossmannek23,Yuan21,Huang23,schuhmacher2024,Zhang22,barison2024,Benfenati21,Mazzola19,McClain24,Fujii22,Schleich23,Takeshita20,Mazzola24}. For example, approaches such as circuit knitting \cite{Bravyi16,Peng20,Mitarai21,Piveteau24,Gian} and entanglement forging \cite{Eddins22,huembeli22,Schoulepnikoff24} enable the simulation of large quantum systems on smaller devices at the expense of a measurement overhead. In quantum embedding methods the system is split into strongly and weakly correlated parts ((in-)active spaces), such that the quantum resources can be utilized to simulate strong interactions while the inactive space is handled classically using, for example, a mean field or a density functional theory approach~\cite{Bauer16,rubin2016,Rossmannek21,Rossmannek23}. Similarly, hybrid quantum-classical wave function ansatze \cite{Yuan21,Huang23,schuhmacher2024,Zhang22,barison2024,Benfenati21,Mazzola19} allow a distribution of resources between a variational quantum circuit and a classical model. However, it is often challenging to choose the optimal partition for a given system, and, in the case of strong interactions, such a partition might not exist at all. Moreover, all hybrid quantum-classical methods currently proposed are formulated in second quantization and are thus subject to basis truncation errors, as stated above.

In this work, we introduce a hybrid method based on variational Monte Carlo \cite{mcmillan65,Becca17} that efficiently combines classical and quantum resources in a global variational ansatz to simulate the ground state of continuous-space systems in first quantization. Our method allows us to harness quantum computers without the need for discretization. As a result, we eliminate basis truncation errors and expensive extrapolations to the infinite basis set.

The variational ansatz is based on parameterized quantum circuits~\cite{Cerezo21b,Benedetti19} adapted to the two considered cases of distinguishable and fermionic particles. In the latter case, we enforce the antisymmetry of the wave function using a Slater determinant of single-particle orbitals. Many-body correlations are captured via so-called backflow transformations~\cite{feynman56,kwon93,taddei15,luo19}, also parameterized by quantum circuits.
Furthermore, we extend the purely quantum ansatz to hybrid quantum-classical ansatze, allowing for a more efficient optimization scheme and improved ground-state energies compared to purely classical wave functions.
Independent of the chosen variational ansatz, the number of qubits on the quantum device depends linearly on the number of particles and the spatial dimension.

We apply our method to three paradigmatic continuous space systems: the one-dimensional quantum rotor model, a gas of Helium-3 particles in one and two spatial dimensions, and the two-dimensional homogeneous electron gas. Note that simulating the latter two systems with conventional techniques is still significantly beyond the reach of quantum devices, considering the current and foreseeable hardware developments~\cite{Shepherd12,Su21}. We show that the accuracy of the ground-state energy, and thus the expressivity of the ansatz, can be controlled by the number of circuit parameters.

For the quantum rotor model we compare our results to matrix product state (MPS) calculations~\cite{white92,schollwock11,orus14} and other typical classical wave function ansatze and demonstrate competing accuracies. We additionally study the effect of shot-noise, which is inevitably present in real quantum hardware experiments.
For the fermionic systems, we compare to the Slater-Jastrow ansatze and show improved accuracy with the quantum circuit ansatz. 
Hence, our framework opens the door to ground state simulations of first-quantized Hamiltonians on small-scale, near-term hardware.

\begin{figure*}[t!]
    \centering
    \includegraphics[width=1.0\textwidth]{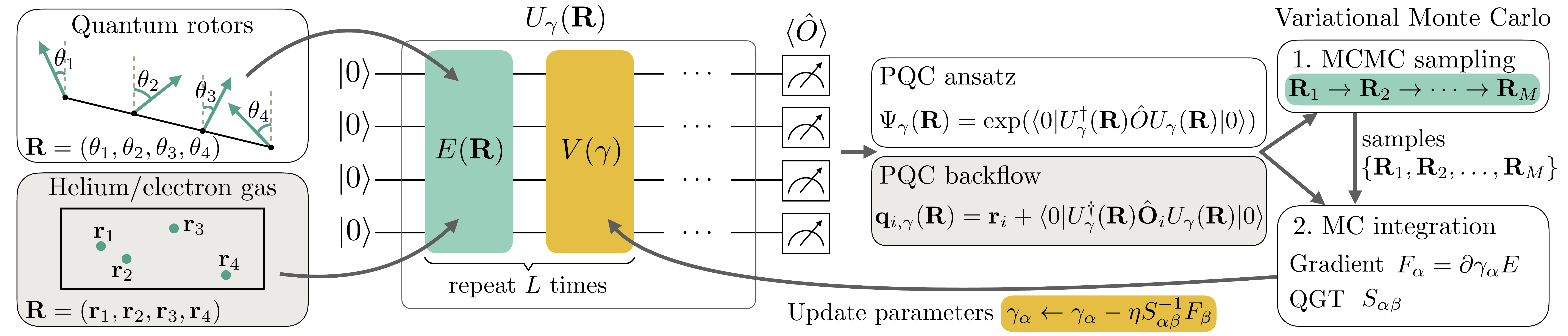}
    \caption{Overview of our framework to find ground states of continuous-space systems (e.g.~quantum rotor model or Helium-3 gas). We represent the variational wave function ansatz in terms of a parameterized quantum circuit (PQC). The circuit unitary $U_\gamma(\mathbf{R})$ depends on optimizable parameters $\gamma$ through a variational unitary $V(\gamma)$, and on the degrees of freedom $\mathbf{R}$ of the system (e.g.~rotor angles or Helium particle positions) through the encoding unitary $E(\mathbf{R})$. The expressivity of the ansatz can be increased by repeating the encoding and variational unitaries. For distinguishable particles (e.g.~quantum rotors) we directly define the log-amplitudes of the wave function as the expectation value of a suitably chosen observable $\hat{O}$ computed on the quantum state. For fermionic systems (e.g.~Helium-3) the wave function is expressed as a classical Slater determinant of single-particle orbitals. To increase expressivity, we modified the single-particle orbitals by backflow transformations, which are again represented as expectation values of the observables $\hat{\mathbf{O}}$. We optimize the ansatze using VMC, i.e., we sample particle configurations from the squared wave function amplitudes via Markov Chain Monte Carlo (MCMC) and use them to compute the gradient of the energy and the quantum geometric tensor (QGT). We then update the parameters and repeat the steps until convergence to the minimal energy is achieved.
    }
    \label{fig:overview}
\end{figure*}

\section{Variational wave functions \label{sec:ansatz}}

Throughout this work we aim to find the ground state $\ket{\Psi_{\mathrm{gs}}}$ of a system of $N$ interacting particles in $d$-dimensional continuous space (e.g. Euclidean space $\mathbb{R}^d$ or the unit sphere $S^d$) governed by a Hamiltonian of the form
\begin{equation}
    H=T+V=-\frac{\hbar^2}{2 m} \sum_{i=1}^N \nabla_{\mathbf{r}_i}^2+V(\mathbf{R}),
\end{equation}
with $T,V$ being the kinetic and potential energies respectively. The ensemble of particle positions are denoted by $\mathbf{R}=(\mathbf{r}_1, \mathbf{r}_2, \dots, \mathbf{r}_N)$.

We approximate $\ket{\Psi_{\mathrm{gs}}}$ by defining a parameterized ansatz state $\Psi_{\gamma}(\mathbf{R})=\braket{\mathbf{R}|\psi_\gamma}$ and subsequently optimize its parameters $\gamma \in \mathbb{R}$ such that the expected value of the energy becomes minimal
\begin{equation}
    \gamma^*=\underset{\gamma}{\arg \min } \ E[\Psi_{\gamma}] = \frac{\braket{\Psi_{\gamma}|H|\Psi_{\gamma}}}{\braket{\Psi_{\gamma}|\Psi_{\gamma}}} .
\end{equation}
The variational principle guarantees that $\min E[\Psi_{\gamma}] \geq E_{\mathrm{gs}}$ and thus, $\ket{\Psi_{\gamma}} = \ket{\Psi_{\mathrm{gs}}}$ iff $E[\Psi_{\gamma}] = E_{\mathrm{gs}}$. The variational principle is exploited in several numerical techniques for ground state optimizations of quantum many-body systems such as the density matrix renormalization group (DMRG) \cite{white92},  variational Monte Carlo (VMC) \cite{mcmillan65,Becca17}, Hartree-Fock~\cite{Helgaker00}, or the variational quantum eigensolver (VQE) \cite{Peruzzo14}. 
In the following, we show how the continuous-space wave function $\Psi_{\gamma}(\mathbf{R})$ can be parameterized using a quantum circuit that respects its general permutation symmetries, and we introduce a scheme to optimize it with a VMC-based approach.

\subsection{Parameterized quantum circuit (PQC) ansatz for distinguishable particles \label{sec:pqc}}

For distinguishable particles, we define the variational ansatz as
\begin{equation}\label{eq:pqc}
    \Psi_{\gamma}(\mathbf{R}) = \exp(\phi_{\gamma}(\mathbf{R}) + i \phi_{\gamma'}(\mathbf{R})),
\end{equation}
where the real and imaginary parts of the log-amplitude, $\phi_{\gamma}(\mathbf{R}),\phi_{\gamma'}(\mathbf{R})\in \mathbb{R}$, are represented by the output of parameterized quantum circuits (PQC), with $\gamma, \gamma'$ the respective circuit parameters. In particular, the log-amplitudes are given by the expectation value of an observable $\hat{O}$ computed on a quantum state $U_\gamma(\mathbf{R})\ket{0}$ that depends on both, the particle coordinates $\mathbf{R}$ and the variational parameters $\gamma$:
\begin{equation}
    \phi_{\gamma}(\mathbf{R}) = \braket{0|U_\gamma^\dagger(\mathbf{R})\hat{O}U_\gamma(\mathbf{R})|0}.
\end{equation}
This expectation value is evaluated by first preparing the quantum state on the quantum computer via the unitary $U_\gamma(\mathbf{R})$ followed by repeated measurements of the qubits.

The exponential form of Eq.~\eqref{eq:pqc} is typically adopted in VMC. Working with a log-space parameterization helps avoid numerical instabilities, since the wave function amplitudes can span several orders of magnitude. Also note that we do not require the ansatz to be normalized.

The expressivity of the resulting ansatz is controlled by the unitary $U_\gamma(\mathbf{R})$. Typically one constructs $U_\gamma(\mathbf{R})$ by repeated applications of an encoding unitary $E(\mathbf{R})$ and a variational unitary $V(\gamma)$ [see Fig.~\ref{fig:overview}]:
\begin{equation}
    U_\gamma(\mathbf{R}) = \prod_{l=1}^L V(\gamma_l) E(\mathbf{R}),
\end{equation}
where $L$ is the total number of layers, i.e., repetitions of the encoding and variational gates. $E(\mathbf{R})$ is responsible for encoding the physical positions of the particles into the quantum state, while $V(\gamma)$ can be optimized to correlate the inputs. Repeating the encoding and variational unitaries enhances the expressivity of the overall ansatz, similarly to increasing the number of hidden layers in a classical neural network \cite{PerezSalinas20,schuld21}. 
To adhere to hardware requirements of current quantum computers, the unitaries typically consist of products of parameterized single-and two-qubit gates. We provide specific examples of such circuits in Section~\ref{sec:models}.

\subsection{Hybrid quantum-classical ansatz \label{sec:hybrid}}

We can further combine classical and quantum resources into a hybrid ansatz by representing the wave function as
\begin{equation}\label{eq:hybrid}
    \Psi_{\gamma}(\mathbf{R}) = \exp(\phi^q_{\gamma}(\mathbf{R}) + \phi^c_{\gamma}(\mathbf{R})), 
\end{equation}
where $\phi^q_{\gamma}(\mathbf{R})$ constitutes the previously discussed PQC model, and $\phi^c_{\gamma}(\mathbf{R})$ is a purely classical ansatz such as a mean-field, Jastrow, or neural network wave function. Note that each model has a distinct set of variational parameters. However, for brevity of notation, we refer to both parameter sets as $\gamma$.

A hybrid quantum-classical ansatz has several advantages: (i) When quantum resources are scarce, we can pre-optimize the classical ansatz $\Psi_{\gamma}(\mathbf{R}) = \exp(\phi^c_{\gamma}(\mathbf{R}))$ independently, to obtain an approximation to the ground state. Afterwards, we add the quantum ansatz, initialized close to $\phi^q_{\gamma}(\mathbf{R})\sim 0$, and continue optimization to reach the desired ground state with a higher accuracy. In this case, the quantum model is used to improve over classical results. (ii) Vice versa, one can start by optimizing the PQC ansatz and add the classical model to improve convergence. This scenario could be useful if the quantum computation is subject to noise and errors or the ansatz has limited expressivity due to hardware limitations, and hence the exact ground state cannot be reached by the PQC alone. In this case, the classical model can be used to improve over the quantum results. (iii) Finally, the hybrid ansatz allows us to use a quantum resource efficient sampling technique which is further discussed in Section~\ref{sec:is}.

\subsection{PQC ansatz for fermionic particles \label{sec:backflow}}
To model indistinguishable fermionic particles, the wave function ansatz is required to be anti-symmetric w.r.t.\ particle permutations. For Mean-Field states (MF), this is typically enforced by using a Slater determinant of single-particle orbitals $\phi_\mu\left(\mathbf{r}_i\right)$ as an ansatz
\begin{align}
    \Psi_{\mathrm{MF}}(\mathbf{R}) =&\Phi_D(\mathbf{R}) \\
    =& \frac{1}{\sqrt{N!}}\left|\begin{array}{cccc}
\phi_1\left(\mathbf{r}_1\right) & \phi_1\left(\mathbf{r}_2\right) & \ldots & \phi_1\left(\mathbf{r}_N\right) \\
\phi_2\left(\mathbf{r}_1\right) & \phi_2\left(\mathbf{r}_2\right) & \ldots & \phi_2\left(\mathbf{r}_N\right) \\
\vdots & \vdots & \vdots & \vdots \\
\phi_N\left(\mathbf{r}_1\right) & \phi_N\left(\mathbf{r}_2\right) & \ldots & \phi_N\left(\mathbf{r}_N\right)
\end{array}\right|
\end{align}
To go beyond the MF form, particle correlations are built into the ansatz by using many-body backflow transformations (BF) \cite{feynman56,kwon93,taddei15,luo19}, i.e., permutation-equivariant coordinate transformations: $\mathbf{Q}(\mathbf{R}) = (\mathbf{q}_{1}(\mathbf{R}), \mathbf{q}_{2}(\mathbf{R}), \dots, \mathbf{q}_{N}(\mathbf{R}))$. $\mathbf{Q}(\cdot)$ takes as input all original particle positions $\mathbf{R}$ and outputs correlated quasiparticle positions $\mathbf{q}_i$. The permutation equivariance condition, i.e.\
\begin{equation}
    \mathbf{q}_j(\mathbf{R}) = \mathbf{q}_i(T_{ij}\mathbf{R}),\quad \mathbf{q}_k(\mathbf{R}) = \mathbf{q}_k(T_{ij}\mathbf{R})\quad \forall i\neq j\neq k, \label{eq:permutation_equivariance}
\end{equation}
where $T_{ij}$ denotes the exchange of particles $i$ and $j$, ensures that the anti-symmetry, enforced by the determinant, is preserved. 
We then replace the original particle positions $\mathbf{r}_i$ with their corresponding quasiparticle positions $\mathbf{q}_i$
\begin{align}
    \Psi_{\mathrm{BF}}(\mathbf{R})& = \Phi_D(\mathbf{Q}(\mathbf{R}))\\
    =  \frac{1}{\sqrt{N!}} & \left|\begin{array}{cccc}
\phi_1\left(\mathbf{q}_1(\mathbf{R})\right) & \phi_1\left(\mathbf{q}_2(\mathbf{R})\right) & \ldots & \phi_1\left(\mathbf{q}_N(\mathbf{R})\right) \\
\phi_2\left(\mathbf{q}_1(\mathbf{R})\right) & \phi_2\left(\mathbf{q}_2(\mathbf{R})\right) & \ldots & \phi_2\left(\mathbf{q}_N(\mathbf{R})\right) \\
\vdots & \vdots & \vdots & \vdots \\
\phi_N\left(\mathbf{q}_1(\mathbf{R})\right) & \phi_N\left(\mathbf{q}_2(\mathbf{R})\right) & \ldots & \phi_N\left(\mathbf{q}_N(\mathbf{R})\right)
\end{array}\right|
\end{align}
It has been shown that the resulting wave function can, in principle, represent any fermionic wave function~\cite{Gabriel22}. In practice, the exact form of the backflow transformation is unknown, and hence found by minimizing the total energy via VMC.

In this work, we parameterize the backflow coordinate transformations via PQCs. Specifically, the particle positions $\mathbf{r}_{i}$ are modified by the expectation value of observables $\hat{\mathbf{O}}_i$ measured on a parameterized quantum state $U_{\gamma}(\mathbf{R})|0\rangle$:
\begin{equation}\label{eq:qbackflow}
    \mathbf{q}_{i,\gamma}(\mathbf{R}) = \mathbf{r}_i + \langle 0|U_{\gamma}^\dagger(\mathbf{R}) \hat{\mathbf{O}}_i U_{\gamma}(\mathbf{R})|0\rangle . 
\end{equation}

To ensure permutation equivariance [Eq.~\eqref{eq:permutation_equivariance}], we require that the observables $\hat{\mathbf{O}}_i$ and the unitary $U_{\gamma}(\mathbf{R})$ in Eq.~\eqref{eq:qbackflow} satisfy $\langle 0|U_{\gamma}^\dagger(\mathbf{R}) \hat{\mathbf{O}}_j U_{\gamma}(\mathbf{R})|0\rangle = \langle 0|U_{\gamma}^\dagger(T_{ij}\mathbf{R}) \hat{\mathbf{O}}_i U_{\gamma}(T_{ij}\mathbf{R})|0\rangle$ [see Appendix~\ref{app:equivariant}]. Specific examples of equivariant PQC ansatze are discussed in Section~\ref{sec:helium}.

Following Eq.~\eqref{eq:hybrid}, we extend this quantum backflow wave function to a hybrid ansatz by multiplying with a classical, permutation-symmetric Jastrow term $J(\mathbf{R})$:
\begin{equation}
    \Psi_{\text{J-BF}}(\mathbf{R}) = J(\mathbf{R}) \Phi_D(\mathbf{Q}(\mathbf{R}))
\end{equation}
In this case, the classical Slater-Jastrow wave function can be pre-optimized and later improved upon by adding the quantum backflow transformations.

\section{Optimization \label{sec:opt}}

\subsection{Variational Monte Carlo (VMC) \label{sec:vmc}}

We optimize the variational wave function ansatz using variational Monte Carlo (VMC) \cite{mcmillan65,Becca17}. 
To that end, we rewrite the expectation values of observables, such as the Hamiltonian, as statistical averages over the normalized probability distribution $\Pi(\mathbf{R}) =|\Psi_\gamma(\mathbf{R})|^2/\int d \mathbf{R}\ |\Psi_\gamma(\mathbf{R})|^2$
\begin{align}
\frac{\langle \Psi_\gamma| H|\Psi_\gamma \rangle}{\braket{\Psi_\gamma|\Psi_\gamma}} &=\int d \mathbf{R}\ \Pi(\mathbf{R}) E_\mathrm{loc}(\mathbf{R}) \equiv \langle E_{\mathrm{loc}}(\mathbf{R}) \rangle_{\Pi}
\end{align}
where $E_{\mathrm{loc}}(\mathbf{R})=-\frac{\hbar^2}{2 m}\sum_i\Psi^{-1}_\gamma(\mathbf{R}) \nabla_{\mathbf{r}_i}^2 \Psi_\gamma(\mathbf{R}) + V(\mathbf{R})$ is the so-called local energy. Monte Carlo integration is used to estimate these expectation values via a set of particle position configurations $\{\mathbf{R}^{[1]},\dots, \mathbf{R}^{[M]}\}$ sampled from the probability distribution $\Pi(\mathbf{R})$:
\begin{align}\langle E_{\mathrm{loc}}(\mathbf{R}) \rangle_{\Pi} \sim \frac 1 M \sum_{j=1}^M E_{\mathrm{loc}}(\mathbf{R}^{[j]}).
\end{align}
In this work, we sample configurations from the target distribution $\Pi(\mathbf{R})$ using conventional Markov Chain Monte Carlo, i.e., using the Metropolis algorithm with a Gaussian proposal distribution. 

To enhance convergence to the ground-state wave function, we employ stochastic reconfiguration (SR) \cite{Sorella98,Becca17}. SR involves estimating the gradient of the energy with respect to each parameter $\gamma_\alpha$
\begin{align}\label{eq:force}
    F_\alpha \equiv  {\partial \gamma_\alpha} E = 2\left(\braket{E_{\mathrm{loc}} O^\dagger_\alpha}_{\Pi} - \braket{E_{\mathrm{loc}}}_{\Pi} \braket{O^\dagger_\alpha}_{\Pi}\right), 
\end{align}
and the quantum geometric tensor (QGT)
\begin{equation}\label{eq:qgt}
    S_{\alpha, \beta}=\left\langle O_\alpha^{\dagger} O_\beta\right\rangle_{\Pi}-\left\langle O_\alpha^{\dagger}\right\rangle_{\Pi}\left\langle O_\beta\right\rangle_{\Pi},
\end{equation}
where $O_\alpha(\mathbf{R})=\partial \gamma_\alpha \log \Psi_\gamma(\mathbf{R})$. Both the gradient and the quantum geometric tensor can be fully expressed in terms of expectation values over the probability distribution $\Pi(\mathbf{R})$ and thus, efficiently approximated via Monte Carlo integration.

The parameter update $\delta \gamma_\alpha$ at each iteration is computed as
\begin{equation}
    \delta \gamma_\alpha = -\eta S^{-1}_{\alpha, \beta} F_{\beta},
\end{equation}
where $\eta$ is the learning rate. Numerical instabilities arising in the inversion of the QGT can be alleviated by adding a small diagonal shift $\epsilon \sim 10^{-3}$ to the QGT, i.e, $S_{\alpha, \beta} + \epsilon \delta_{\alpha, \beta}$.

Note that the SR approach is formally equivalent to performing the shifted power method in the variational manifold, and reduces to imaginary time evolution in the limit of continuous updates ($\eta\rightarrow 0)$ \cite{Becca17}. 

\subsection{VMC with PQC ansatz \label{sec:qvmc}}

Given a set of samples $\{\mathbf{R}^{[1]},\dots, \mathbf{R}^{[M]}\}\sim\Pi(\mathbf{R}) $, we need to evaluate the energy gradient $F_\alpha$ and the quantum geometric tensor $S_{\alpha\beta}$ to update the parameters $\gamma_\alpha$. This requires us to compute $E_{\mathrm{loc}}(\mathbf{R}) = T_{\mathrm{loc}}(\mathbf{R}) + V_{\mathrm{loc}}(\mathbf{R})$ and $O_\alpha(\mathbf{R})=\partial \gamma_\alpha \log \Psi_\gamma(\mathbf{R})$ for each sample. Using $\Psi_{\gamma}(\mathbf{R}) = \exp(\phi_{\gamma}(\mathbf{R}))$ as the ansatz, the local energy $T_{\mathrm{loc}}(\mathbf{R})$ becomes
\begin{align}
T_{\mathrm{loc}}(\mathbf{R})=-\frac{\hbar^2}{2 m} \sum_i \left(\nabla_{\mathbf{r}_i} \phi_{\gamma}(\mathbf{R})\right)^2 + \left(\nabla_{\mathbf{r}_i}^2\phi_{\gamma}(\mathbf{R})\right).
\end{align}
If the log-amplitudes, $\phi_{\gamma}$, are given by expectation values of PQCs, i.e., $\phi_{\gamma}(\mathbf{R})=\braket{0|U_\gamma^\dagger(\mathbf{R})\hat{O}U_\gamma(\mathbf{R})|0}$, we compute their first- and second-order derivatives using a (generalized) parameter shift rule \cite{Mitarai18,Schuld19,Wierichs2022} [see Appendix~\ref{app:param_shift} for details]. Similarly, the derivative with respect to the variational parameters, $O_\alpha(\mathbf{R})=\partial \gamma_\alpha \phi_{\gamma}(\mathbf{R})$, can be expressed as a parameter shift rule as long as the parameters enter via angles of (multi-qubit) rotation gates.

For hybrid quantum-classical ansatze or quantum backflow transformations, we instead use the chain and product rules to simplify the derivatives. Gradients of parameters entering in the unitaries of the quantum circuit are again computed via parameter shift rules; derivatives of classical parameters are calculated via conventional automatic differentiation. In this work, we use the Python libraries \texttt{Pennylane} \cite{pennylane} for obtaining gradients on quantum circuits and \texttt{NetKet} \cite{netket3,netket2} for an implementation of VMC. The full algorithm underlying the optimization is outlined in Algorithm~\ref{alg} in Appendix~\ref{app:vmc}. The code is available on \texttt{GitHub}~\cite{github}.

\subsection{Quantum resource-efficient VMC \label{sec:is}} 
As previously described, the samples of the particle position configurations are generated using Markov Chain Monte Carlo. Each Metropolis step involves evaluating the amplitudes of the wave function $\Psi_\gamma(\mathbf{R}')$ in the candidate configuration $\mathbf{R}'$ to calculate the acceptance ratio. Thus, for every step of the chain, we have to compute the respective expectation values on the PQC ansatz. Overall, the number of resulting circuit evaluations can become high as (i) not all candidate positions $\mathbf{R}'$ are accepted, (ii) a number of subsequent samples are usually correlated and have to be discarded, and (iii) the first few samples of the Markov chain generally don't follow the correct distribution and thus have to be discarded.

To reduce the use of quantum resources in our framework, we introduce a quantum resource-efficient VMC scheme that is applicable to hybrid quantum-classical ansatze of the form $\Psi_{\gamma}(\mathbf{R}) = \psi^q_{\gamma}(\mathbf{R}) \psi^c_{\gamma}(\mathbf{R})$ and does not require any circuit evaluations for the sampling subroutine. We achieve this by using importance sampling, which allows us to sample from the classical distribution $\Pi_c(\mathbf{R})=|\psi^c_\gamma(\mathbf{R})|^2/\int d \mathbf{R}\ |\psi^c_\gamma(\mathbf{R})|^2$ only, rather than the target probability distribution $\Pi(\mathbf{R})$. To correct for any deviations between the distributions, we include an additional weight factor $w^q_{\gamma}(\mathbf{R}) = |\psi^q_{\gamma}(\mathbf{R})|^2 / \braket{|\psi^q_{\gamma}(\mathbf{R})|^2}_{\Pi_c}$ in the computations of all expected values. Thus, the energy gradient becomes
\begin{align}
    F_\alpha =  2\left(\braket{w^q_{\gamma}E_{\mathrm{loc}} O^\dagger_\alpha}_{\Pi_c} - \braket{w^q_{\gamma}E_{\mathrm{loc}}}_{\Pi_c} \braket{w^q_{\gamma}O^\dagger_\alpha}_{\Pi_c}\right), 
\end{align}
and similarly for the quantum geometric tensor Eq.~\eqref{eq:qgt}.

Importance sampling can lead to an increase in the variance of the quantities to be estimated. To counteract this effect, we require a larger number of samples for the Monte Carlo integration which can render the scheme inefficient. However, the variance is expected to remain small as long as $w^q_{\gamma}(\mathbf{R}) \approx 1$. 
Therefore, we propose to first pre-optimize the classical wave function ansatz $\psi^c_{\gamma}(\mathbf{R})$ to obtain a good approximation to the ground state. The PQC wave function is then initialized to unity and regarded as a small correction such that we have $w^q_{\gamma}(\mathbf{R}) \approx 1$, throughout the optimization.

\section{Results \label{sec:models}}
\subsection{Quantum rotor model \label{sec:rotor}}

\begin{figure}[t!]
    \centering
    \includegraphics[width=0.48\textwidth, left]{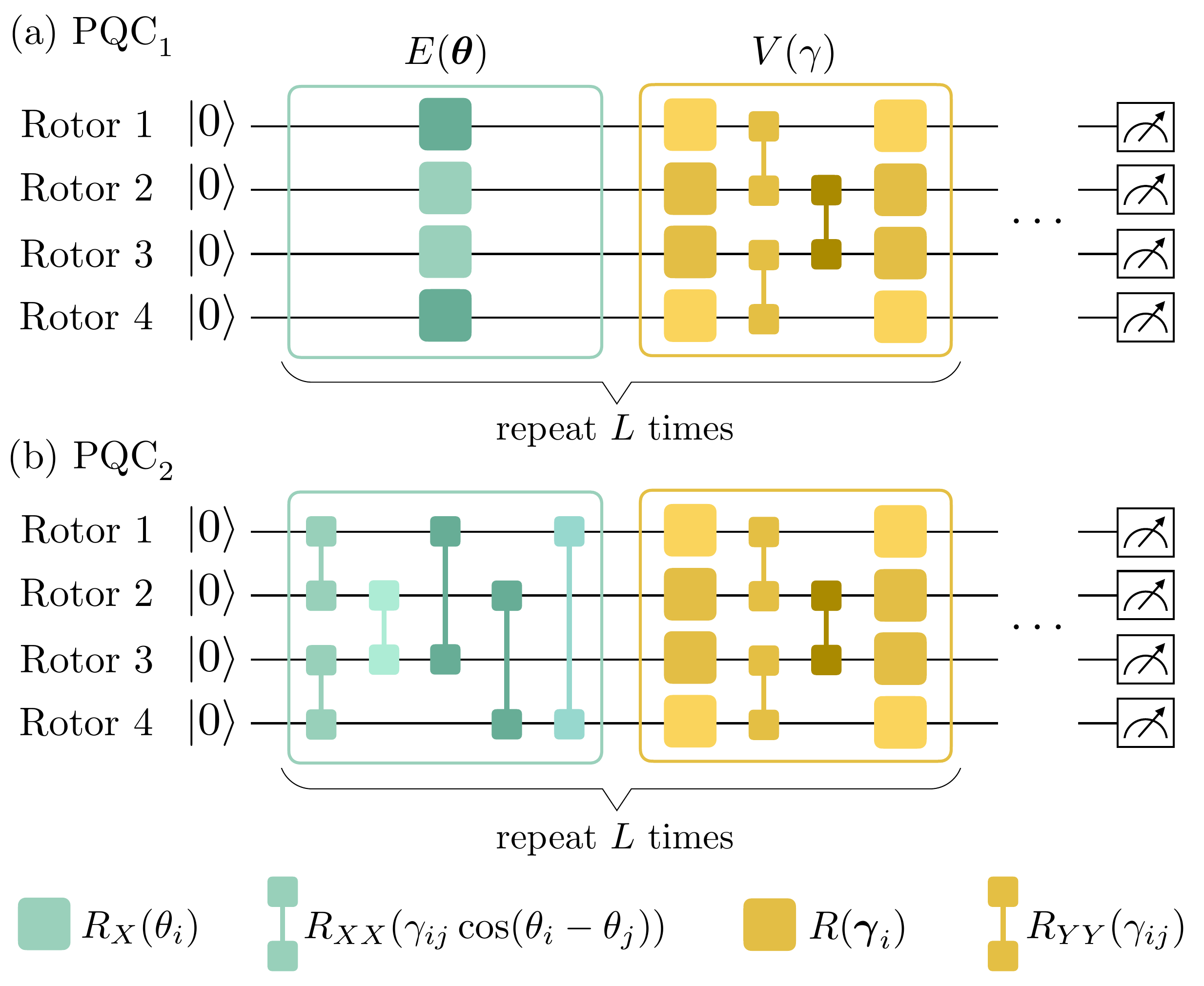}
    \caption{Quantum rotor model --- PQC ansatze consisting of $L$ repeated layers of alternating applications of an encoding unitary $E(\boldsymbol{\theta})$ and a variational unitary $V(\gamma)$. The variational unitary is comprised of two layers of arbitrary single-qubit rotations parameterized by three angles and a hardware-efficient layer of two-qubit $R_{YY}$ gates. {\bf (a)} In PQC$_1$ the rotor angles $\theta_i$ are encoded via the angles of single-qubit rotation $R_X$ gates. {\bf (b)} In PQC$_2$ the two-body features $\cos(\theta_i - \theta_j)$ between rotors $i$ and $j$ are encoded via the angles of two-qubit rotation $R_{XX}$ gates acting on the respective qubits (rotors). We impose reflection symmetry across the center of the chain by fixing the corresponding parameters in a given layer to be the same (indicated by different color shadings).
    }
    \label{fig:rotor_circ}
\end{figure}

As a first example, we consider the one dimensional quantum rotor model on a chain [see Fig.~\ref{fig:overview}] which can be used to model bosons in optical lattices \cite{cha91} or arrays of superconducting Josephson junctions \cite{Frisk19}. Each rotor is confined to the unit circle and characterized by a single angle $\theta_i\in[0,2\pi], i=1,\dots ,N$. 
We consider open boundary conditions and nearest neighbor interactions only. Thus, the corresponding Hamiltonian is given by
\begin{equation}\label{eq:rotor}
    H=-\frac{1}{2} \sum_{i=1}^N \frac{\partial^2}{\partial \theta_i^2}- \sum_{i=1}^{N-1} \cos \left(\theta_i-\theta_{i+1}\right),
\end{equation}
where the mass and interaction strength have been set to unity, and we work in units where $\hbar=1$. 

\begin{figure*}[t!]
    \centering
    \includegraphics[width=1.0\textwidth]{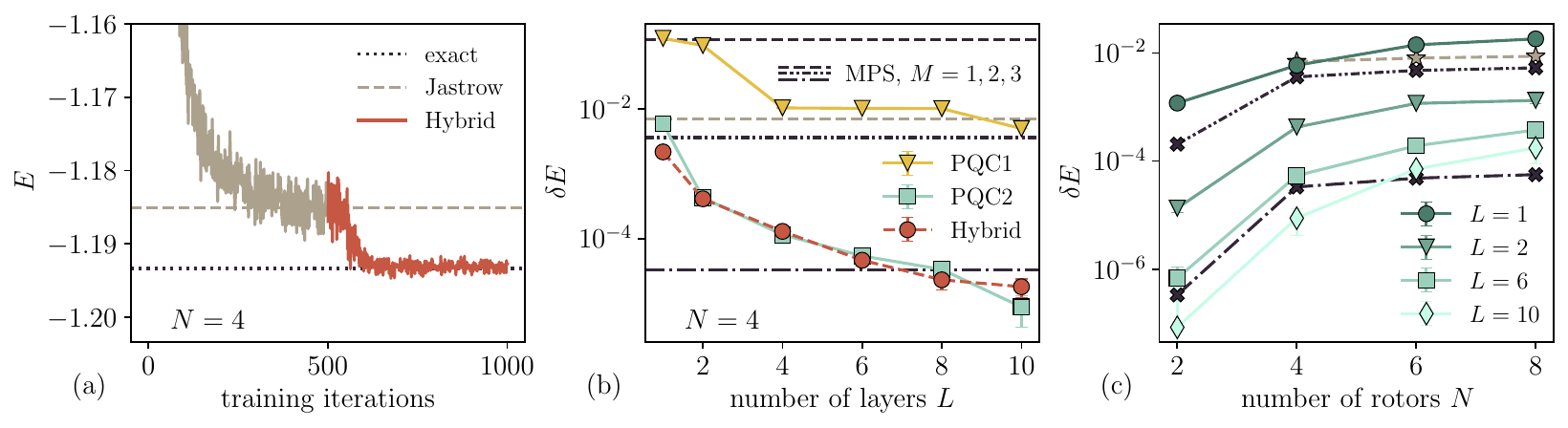}
    \caption{Quantum rotor model --- {\bf (a)} Training curve for the hybrid quantum-classical ansatz with $N=4$ rotors employing noise-free statevector simulations. The estimated energy at each training iteration is shown together with the exact ground state energy (black dotted line). For the first 500 iterations a classical Jastrow ansatz is optimized (gray). Afterwards, a PQC$_2$ with $L=4$ layers is introduced, initialized to unity, and optimized further (red). The hybrid ansatz is able to quickly improve on the pre-optimized Jastrow wave function achieving energies close to the exact ground state energy. {\bf (b)} The relative error to the exact ground state energy $\delta E = |(E-E_{\text{exact}}) / E_{\text{exact}}|$ as a function of the number of layers for different PQC ansatz choices. The PQC$_1$ (single-qubit encoding) has the lowest expressivity, only reaching accuracies comparable to the classical Jastrow (gray dashed line). The PQC$_2$ (two-qubit uploading) and the hybrid ansatz (PQC$_2$ + Jastrow) yield similar energies that can be systematically improved by adding more circuit layers. Black dotted lines indicate the accuracy obtained via MPS simulations of the discretized rotor model with varying Hilbert space cutoffs $M$. Error bars denote the estimated standard deviation arising from the finite number of samples used for the Monte Carlo integration. {\bf (c)} The relative error to the exact ground state energy attained with PQC$_2$ for different system sizes $N$ and number of layers $L$. Jastrow (gray stars) and MPS energies (black crosses) are shown as a benchmark.
    }
    \label{fig:rotor1}
\end{figure*}

A classical VMC ansatz for the rotor model is the Jastrow wave function, which will also serve as a benchmark in the following
\begin{equation}
    \Psi_{\text{Jas}}(\boldsymbol{\theta}) = \exp\left(\sum_{n=1}^{n_{\text{max}}}\sum_{i=1}^{N-n}\sum_{k=1}^{k_{\text{max}}} \!c_{n,i,k}\cos(k(\theta_i-\theta_{i+n}))\right).
\end{equation}
In the expression above, the index $n$ determines the interaction range, $i$ indexes the rotors in the chain, $k$ corresponds to the frequency of the cosine series, and the coefficients $c_{n,i,k}$ denote the variational parameters. Note that for $N=2$ rotors the Jastrow wave function exactly represents the true ground state wave function in the limit of $k_{\text{max}} \rightarrow \infty$. However, for $N>2$ rotor systems the Jastrow ansatz is not exact and more expressive parameterizations have to be considered such as neural networks \cite{stokes23,Matija23}.

We use $\Psi_{\text{PQC}}(\boldsymbol{\theta}) = \exp(\braket{0|U_\gamma^\dagger(\boldsymbol{\theta})\hat{O}U_\gamma(\boldsymbol{\theta})|0})$ as the quantum ansatz. The circuit comprises $N$ qubits such that each qubit can be associated with one rotor in the chain [see Fig.~\ref{fig:rotor_circ}]. 
The encoding unitary at layer $l$ is chosen as
\begin{equation}
    E^{(l)}(\boldsymbol{\theta}) = \prod_{i<j} R_{X_i X_j}(\gamma^{(l)}_{ij}\cos(\theta_i-\theta_j)),
\end{equation}
where $R_{X_i X_j}(\alpha) = \exp(-i \frac{\alpha}{2} X_i X_j)$ is a two-qubit rotation gate applied to the qubits (rotors) $i$ and $j$. Thus, we effectively encode the two-body interactions between each rotor pair into the quantum circuit. Note that we also include variational parameters, $\gamma^{(l)}_{ij}$, in the encoding unitary which increases the overall expressivity of the ansatz \footnote{It has been shown that PQCs give rise to a Fourier series in the uploaded data. The frequency of this series is determined by the spectrum of the encoding unitary generators. Without additional parameters in the encoding, the available frequencies are fixed and their number only scales linearly with number of circuit layers. However, by introducing variational parameters into the encoding, and thus, modifying the spectrum of each generator individually, the resultant model can, in principle, harness an exponentially large space of frequencies.}. We compare this two-qubit embedding approach (referred to as PQC$_2$) with a simple single-qubit encoding scheme (PQC$_1$) where the bare rotor angles are directly uploaded as the angles of single-qubit rotation gates, i.e., $E^{(l)}(\boldsymbol{\theta}) = \prod_{i} R_{X_i}(\theta_i)$, while keeping the variational circuit the same [see Fig.~\ref{fig:rotor_circ}(a)].

The variational unitary is given by
\begin{align}
    V^{(l)} = \Bigg[ \prod_{i=1}^{N} R_i(\gamma^{(l)}_{1,i},\gamma^{(l)}_{2,i},\gamma^{(l)}_{3,i}) \prod_{i=1}^{N-1} R_{Y_i Y_{i+1}}(\gamma^{(l)}_{4,i})\\ 
    \prod_{i=1}^{N} R_i(\gamma^{(l)}_{5,i},\gamma^{(l)}_{6,i},\gamma^{(l)}_{7,i}) \Bigg],
\end{align}
and composed of two layers of arbitrary single-qubit rotations $R_i$ parameterized by three Euler angles, and a layer of parameterized two-qubit gates applied to nearest-neighbor qubits only.

The parameters $\gamma$ are different for every layer $l$. However, we enforce reflection symmetry across the center of the chain by fixing the respective parameters within each layer to be the same [see Fig~\ref{fig:rotor_circ}].  Finally, we choose a weighted sum of Pauli-$\hat{Z}$ operators $\hat{O} = \sum_i c_i \hat{Z}_i$ as the observable to measure, where the coefficients $c_i$ represent additional, fully classical and trainable parameters~\footnote{The measurement of observables other than Pauli-$Z$ amounts to performing a basis change, which can be absorbed into the variational unitary $U_\gamma(\mathbf{R})$ and thus, in principle, be learned.}.

We first consider the case of $N=4$ rotors and optimize the wave function ansatze with the single-qubit (PQC$_1$) and the two-qubit (PQC$_2$) encodings, using noise-free statevector simulations. Fig.~\ref{fig:rotor1}(b) shows the achieved relative error to the exact ground-state energy, $\delta E = |(E-E_{\text{exact}}) / E_{\text{exact}}|$, as a function of the number of layers $L$. As a benchmark, we also plot the energies obtained from optimizing a Jastrow ansatz (gray dashed line) and from MPS calculations with varying Hilbert space cutoffs $M$ (black dashed lines). The PQC$_1$ achieves a comparable accuracy to the Jastrow wave function for $L\geq 4$. However, increasing the number of layers does not provide systematically improved accuracies, suggesting that this ansatz choice has only limited expressivity. In contrast, PQC$_2$ outperforms the Jastrow ansatz already for $L=2$ layers and the accuracy further improves with increasing circuit depth. For example, the PQC$_2$ with $L=8$ layers yields results comparable to an exact MPS simulation with a local Hilbert space cutoff $M=3$. Simulating this discrete system on a quantum computer would require at least 12 qubits instead of the 4 qubits considered here [see Appendix~\ref{app:discrete}].

The enhanced performance of PQC$_2$ over PQC$_1$ can be attributed to the correlated embedding of the rotor angle features and the additional parameters in the encoding unitaries. The latter allow the parameterized circuit to represent a larger space of Fourier series with variable (learnable) frequencies \cite{schuld21,jaderberg24}. On the other hand, in the simple PQC$_1$ encoding the Fourier frequencies are not learnt. In Appendix~\ref{app:comparison} we examine the expressivity of different coupling maps (e.g.~nearest-neighbor or all-to-all) for the variational and encoding unitaries in the PQC$_2$ circuit and study their effects on the final precision of the simulation.

Next, we turn to hybrid quantum-classical models and represent the wave function as a product of the classical Jastrow and the PQC$_2$ ansatz $\Psi_{\gamma}(\boldsymbol{\theta}) = \Psi_{\text{Jas}}(\boldsymbol{\theta})\Psi_{\text{PQC}}(\boldsymbol{\theta})$. We pre-optimize the Jastrow factor using standard VMC. Once it is converged, the PQC ansatz is added and initialized close to unity. The subsequent optimization is carried out using the quantum resource-efficient VMC, i.e., we only sample from the optimized classical Jastrow wave function. The resulting learning curve of the two-stage training is depicted in Fig.~\ref{fig:rotor1}(a). The hybrid ansatz is able to improve on the pre-optimized Jastrow wave function and quickly converges to the ground state. Fig.~\ref{fig:rotor1}(b) shows the achieved accuracy of the hybrid ansatz (red dashed line) for different number of layers.  We observe comparable results as obtained with the PQC$_2$ ansatz alone. However, the hybrid approach requires fewer quantum circuit evaluations as the Monte Carlo sampling step is fully classical.

To study the scaling of our method with system size, we repeat the simulations with the PQC$_2$ ansatz for different numbers of rotors $N$ and plot the achieved relative errors in Fig.~\ref{fig:rotor1}(c). As a benchmark, we also show the results from the discretized MPS simulations with Hilbert space cutoffs $M=2,3$. As expected the errors in energy increase as more rotors are added to the chain. Hence, to maintain a certain accuracy across different system sizes, we need to increase the number of layers of the ansatz circuit. In contrast, to improve the accuracy of a discretized simulation we would inevitably also require additional qubits. Thus, our method can be viewed as a means of trading a larger number of qubits for an increased circuit depth and additional circuit evaluations.

\begin{figure}[t!]
    \centering
    \includegraphics[width=0.4\textwidth]{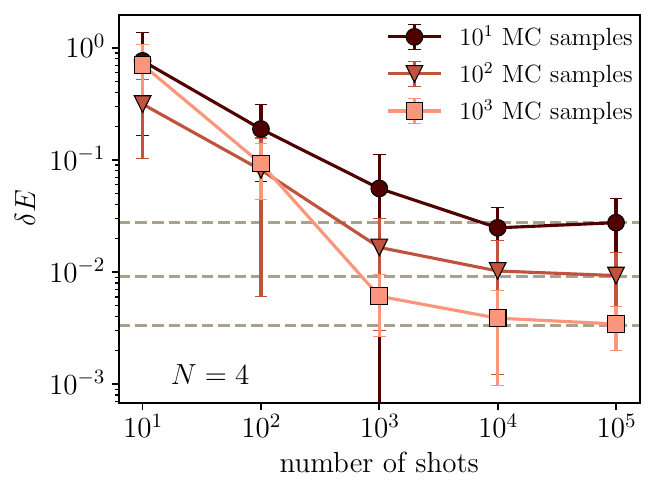}
    \caption{Quantum rotor model --- Testing the performance of the optimized PQC$_2$ ansatz with $L=1$ layers in the presence of measurement shot noise. The relative error to the exact ground state energy is plotted against the number of shots used per expectation value computation. Different colors indicate the number of Monte Carlo (MC) samples that are used for estimating the energies. Dashed lines show the results from statevector simulations in which expectation values of the quantum circuit are calculated exactly.
    }
    \label{fig:rotor2}
\end{figure}

Finally, we analyze the effects of shot noise that is caused by the finite number of measurements to  evaluate expectation values on quantum devices. To that end, we calculate the energy of the already optimized PQC$_2$ wave function with $L=1$ layers as a function of shots per expectation value estimation and number of Monte Carlo samples. Fig.~\ref{fig:rotor2} displays the average achieved error to the exact ground state energy (each line corresponds to a different number of Monte Carlo samples). As expected, the accuracy can be enhanced both, by increasing the number of Monte Carlo samples or the number of shots per expectation value. Note that the total number of required circuit evaluations scales with the product of the number of shots and the number of samples and the optimal distribution of resources is not immediately evident. For example, given a fixed budget of circuit evaluations, Fig.~\ref{fig:rotor2} suggests that beyond $10^3$ shots it is favorable to increase the number of samples whereas the opposite is the case in other regimes. A more detailed analysis of the effects of sampling noise are provided in Appendix~\ref{app:noise}.

\subsection{Helium-3 \label{sec:helium}}

\begin{figure}[t!]
    \centering
    \includegraphics[width=0.48\textwidth, left]{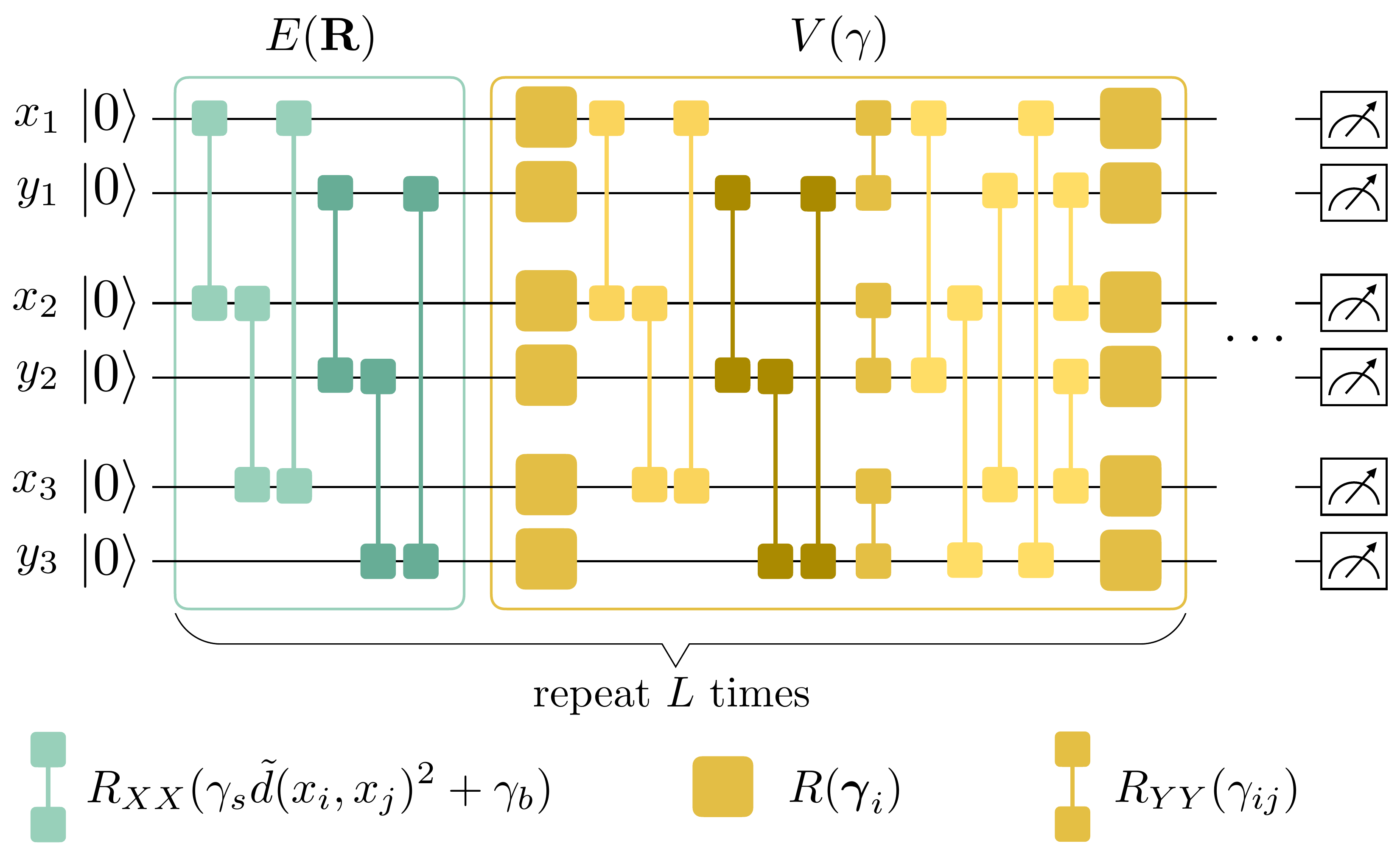}
    \caption{Helium-3 --- PQC ansatz used for the quantum backflow transformation in $d=2$ dimensions. Two qubits are associated to every particle, one qubit for each dimension. The periodic surrogate distances $\tilde{d}\left(x_i, x_j\right)$, $\tilde{d}\left(y_i, y_j\right)$ between different particle coordinates are encoded as the angles of two-qubit rotation $R_{XX}$ gates acting on the respective qubits. The variational unitary consists of two layers of parameterized single-qubit rotations $R$ and a layer of parameterized two-qubit $R_{YY}$ gates connecting qubits associated to different particles and/or dimensions. The distinct color shadings indicate which parameters are shared between two-qubit gates such that the backflow transformation is equivariant with respect to particle exchanges.
    }
    \label{fig:helium_circ}
\end{figure}

To showcase the applicability of our approach to fermionic models, we consider a system of $^3$He particles in one and two spatial dimensions subject to periodic boundary conditions. The effective interactions of the Helium particles are modeled by the Aziz potential \cite{aziz77} which resembles a Lennard-Jones hard-core potential for short distances and quickly decays at large distances [see Appendix~\ref{app:aziz} for its exact form]. The full Hamiltonian is defined as
\begin{equation}\label{eq:helium}
H=-\frac{1}{2} \sum_{i=1}^N \nabla_{\mathbf{r}_i}^2+  \sum_{i<j} V_{\text {Aziz }}\!(d\left(\mathbf{r}_i, \mathbf{r}_j\right)),
\end{equation}
where $d\left(\mathbf{r}_i, \mathbf{r}_j\right) = \left\|\mathbf{r}_i - \mathbf{r}_j\right\|$ is the Euclidean distance between Helium particles which is calculated using the minimum image convention to adhere to the periodic boundary conditions \cite{Allen04}.

We consider $N\!=\!3$ Helium particles and set the density to $\rho \!=\! 0.3 \,\text{\r{A}}^{-1}$ in $d\!=\!1$ dimensions and to $\rho \!=\! 0.03 \,\text{\r{A}}^{-2}$ for $d\!=\!2$. To benchmark our results we optimize a classical Slater-Jastrow wave function ansatz $\Psi_{\text{S-J}}(\mathbf{R}) = J(\mathbf{R}) \Phi_D(\mathbf{R})$ where the symmetric Jastrow factor is given by
\begin{equation}
    J(\mathbf{R}) = \prod_{i<j} \exp \left[-\frac{c}{2} \tilde{d}\left(\mathbf{r}_i, \mathbf{r}_j\right)^{-5}\right],
\end{equation}
with $c$ being a trainable parameter and $\tilde{d}\left(\mathbf{r}_i, \mathbf{r}_j\right) = \left\|\frac{L}{2} \sin \left(\frac{\pi}{L} (\mathbf{r}_i - \mathbf{r}_j)\right)\right\|$ representing a periodic surrogate of the Euclidean distance. $L$ is the size of the simulation cell. The form of the Jastrow is imposed by Kato’s cusp condition \cite{kato57} which ensures that the energy stays finite at short inter-particle distances where the potential diverges. Finally, for the Slater determinant $\Phi_D(\mathbf{R})$ we choose plane waves as single-particle orbitals $\phi_\mu\left(\mathbf{r}_i\right) = \exp(i \mathbf{k}_\mu \cdot \mathbf{r}_i)$ with $\mathbf{k}_\mu=\frac{2 \pi}{L} \mathbf{n}, \mathbf{n} \in \mathbb{Z}^d$.

To increase the expressivity of the simple Slater-Jastrow ansatz we add a backflow transformation that is parameterized by a PQC [see Section~\ref{sec:backflow}]. In the case of one spatial dimension, we can again associate each helium particle with one qubit in the circuit. The encoding unitary has a similar structure and elementary gates as in the previously discussed rotor model example [cf.~Fig.~\ref{fig:rotor_circ}(b)]. Specifically, we have $E^{(l)}(\mathbf{R}) = \prod_{i<j} R_{X_i X_j}(\gamma_s^{(l)}\tilde{d}\left(x_i, x_j\right)^2 + \gamma_b^{(l)})$. The scale $\gamma_s^{(l)}$ and bias $\gamma_b^{(l)}$ are optimizable parameters that are each shared across all qubits per layer in order to fulfil the equivariance condition of the backflow transformation. The variational unitary is chosen as $V^{(l)} = \prod_{i} R_i(\boldsymbol{\gamma}^{(l)}_{1}) \prod_{i<j}  R_{Y_i Y_{j}}(\gamma^{(l)}_{2}) \prod_{i} R_i(\boldsymbol{\gamma}^{(l)}_{3})$, where parameters are again shared within a layer. Furthermore, we require that all two-qubit gates are applied to every pair of qubits to obtain a permutation equivariant circuit ansatz.

Finally, we define the backflow-transformed coordinate of quasiparticle $i$ in terms of the Pauli-$\hat{Z}$ expectation value measured on qubit $i$
\begin{align}\label{eq:hel1d}
    q_{i,\gamma}(\mathbf{R}) =  x_i + c_{\mathrm{real}} \langle \hat{Z}_i \rangle_{\mathbf{R}, \gamma_\mathrm{real}} + i c_\mathrm{imag} \langle \hat{Z}_i \rangle_{\mathbf{R}, \gamma_\mathrm{imag}} ,
\end{align}
where we defined $\langle \hat{Z}_i \rangle_{\mathbf{R}, \gamma}=\braket{0|U_\gamma^\dagger(\mathbf{R})\hat{Z}_i U_\gamma(\mathbf{R})|0}$. Consequently, we train two circuits with different parameter sets $\gamma_\mathrm{real}$ and $\gamma_\mathrm{imag}$ and add their respective outputs as the real and imaginary parts to the original particle positions, which allows the backflow transformation to modify both the real and imaginary parts of the single-particle orbitals independently. The coefficients $c_{\mathrm{real}}$ and $c_{\mathrm{imag}}$ in the expression above represent additional classical parameters.

For $d>1$ dimensions, we can employ quantum circuits with $dN$ qubits and associate a different set of $d$ qubits to each particle. For the encoding, we choose $E^{(l)}(\mathbf{R}) = \prod_{i<j} \prod_{r=1}^d R_{X_{ir} X_{jr}}(\gamma_s^{(l)}\tilde{d}\left(x_{ir}, x_{jr}\right)^2+ \gamma_b^{(l)}))$
where $i,j$ indexes particles, and $r$ the spatial dimension of the Cartesian coordinates [see Fig.~\ref{fig:helium_circ}]. Similarly, we adjust the variational unitary to include additional two-qubit gates that connect qubits associated to different spatial dimensions. We then generalize the definition of the 1d backflow transformation to $q_{ir,\gamma}(\mathbf{R}) =  x_{ir} + c_\mathrm{real} \langle \hat{Z}_{ir} \rangle_{\mathbf{R}, \gamma_\mathrm{real}} + i c_\mathrm{imag} \langle \hat{Z}_{ir} \rangle_{\mathbf{R}, \gamma_\mathrm{imag}}$. We show in Appendix~\ref{app:equivariant} that the obtained backflow transformation is indeed equivariant with respect to particle transpositions.

\begin{figure}[t!]
    \centering
    \includegraphics[width=0.45\textwidth, left]{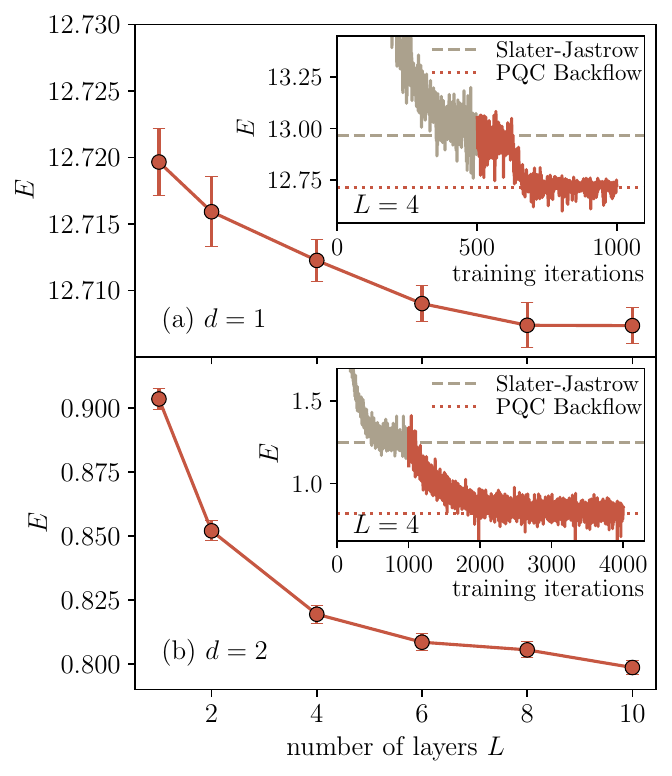}
    \caption{Helium-3 --- Achieved energies of the PQC backflow ansatz as a function of the number of circuit layers for a $d=1$ {\bf (a)} and $d=2$ {\bf (b)}  dimensional system using noise-free statevector simulation. Error bars indicate the estimated standard deviation originating from the stochastic Monte Carlo integration. Insets:~Learning curves for the case of $L=4$ layers. For the first 500 (1000) iterations a classical Slater-Jastrow ansatz is pre-optimized (gray). Upon convergence, the PQC backflow transformation is added, initialized to zero, and optimized (red). The quantum backflow ansatz is able to lower the energy further, beating the classical Slater-Jastrow wave function.
    }
    \label{fig:helium}
\end{figure}

We again use the hybrid quantum-classical scheme and pre-optimize the classical Slater-Jastrow ansatz using VMC. Once converged, we replace the Helium particle positions $\mathbf{r}_i$ in the Slater determinant with the quantum backflow transformed coordinates $\mathbf{q}_{i,\gamma}(\mathbf{R})$ and continue training. Representative learning curves for the 1d and 2d cases are shown as insets in Fig.~\ref{fig:helium}. In both instances, the quantum backflow transformation is able to improve on the classical Slater-Jastrow ansatz reaching lower energies. Similarly to the previous rotor model example, we can systematically reduce the variational energies by increasing the number of layers, and thus the expressivity of the PQC ansatz [see Fig.~\ref{fig:helium}]. For $d=1$, the energy converges already for a small number of layers as we observe only small improvements for $L>1$. In contrast, the ground state for the $d=2$ system is more difficult to represent and learn, requiring a larger number of optimization iterations and deeper variational circuits.

\subsection{Difficulty of simulating continuous-space systems with existing quantum algorithms}

In the following we will estimate the quantum resource requirements of existing quantum algorithms and compare them to our approach, showcasing the merit of the hybrid method for near- and intermediate-term quantum hardware.

Simulating continuous space systems using existing quantum algorithms requires discretization of the system's state space, such that it can be mapped to qubits. Following a second quantized approach, this amounts to choosing a finite set of basis functions, which then dictates the computational resources required. 
However, finding a suitable set of basis functions for a given problem is not always feasible. For example for the Hamiltonian of Eq.~\eqref{eq:helium}, plane waves are a natural choice due to the periodic boundary conditions. However, additional constraints on the wave function, such as an asymptotic decay to zero amplitude for vanishing distance between particles caused by a diverging interaction energy, are difficult to fulfill using plane waves. In fact, we found that a prohibitively large number of plane-wave orbitals is required to faithfully represent this decay.
Even though there might be more suitable basis sets for this particular problem, the above stresses the general difficulty of simulating continuous-space systems on near-term quantum computers using discretization approaches. In contrast, we can embed the necessary constraints directly into the ansatz, eliminating all issues arising from a bad choice of basis functions.

Similarly, simulating the homogeneous electron gas (HEG) [Appendix~\ref{app:heg}] in a discrete basis on a quantum device would require a significantly larger number of qubits than the $dN$ qubits needed by our scheme. 
For example, employing the classical full configuration interaction quantum Monte Carlo (FCIQMC) approach, Shepherd et al.~\cite{Shepherd12} performed simulations with up to 2838 orbitals in order to extrapolate the energy of a $N=54$ electron gas to the continuum limit. Hence, a quantum simulation reaching similar accuracy would require 2838 qubits whereas our scheme would involve only 162 qubits, a nearly 18-fold reduction. This discrepancy is expected to be further enhanced for larger HEG interactions than the ones considered by Shepherd et al.~\cite{Shepherd12}. In terms of fault-tolerant quantum algorithms operating in first quantization, Su et al.~\cite{Su21} estimated that roughly 2000 logical (error corrected) qubits are required to simulate a $N=50$ HEG within chemical accuracy using a method based on qubitization. Thus, simulations of the electron gas with conventional, non-variational techniques will be out-of-reach for near-term devices featuring only a handful of error-corrected qubits. We therefore expect our framework to be especially beneficial in the interim period until large-scale fault-tolerant quantum computers have been realized.

\section{Conclusion \& Outlook\label{sec:conclusion}}

In this work, we proposed a VMC based method for finding ground states of continuous-space systems that leverages quantum resources for parameterizing the wave function ansatz. We presented two general choices of ansatze:~(i)~Representing the log-amplitudes of the wave function as an expectation value computed on a parameterized quantum circuit. This ansatz can be applied to systems composed of distinguishable particles. (ii)~For fermionic systems we employed a classical Slater determinant of single-particle orbitals in combination with parameterized backflow transformations, represented as expectation values computed on variational quantum circuits. Finally, we also provided examples of hybrid quantum-classical ansatze with a scheme for reducing the required quantum resources for optimization.

We applied our method to the quantum rotor model [Sec.~\ref{sec:rotor}], a system of Helium-3 particles in one and two dimensions [Sec.~\ref{sec:helium}], and the homogeneous electron gas [App.~\ref{app:heg}]. We studied how the accuracy in the ground-state energy depends on the choice of ansatz, the number of circuit layers, and the size of the system. In particular, we demonstrated that if the encoding of particle position features into the quantum circuit is sufficiently expressive, the precision of the ground-state energy can be tuned by adjusting the number of circuit parameters. Finally, using shot-based simulations, we analyzed the effect of sampling noise on the final results.

Our approach avoids errors introduced by discretization. Furthermore, the required number of qubits in our scheme is fixed by and scales linearly with the total number of particles, in contrast to discretization based methods, where the qubit count depends on the accuracy of the discretization. For many continuous-space problems of interest, in particular the electronic structure of molecules and materials, the required number of qubits is still significantly out of reach for near-term quantum computers \cite{Su21,Burg21,Beverland22,Reiher17,Babbush18}, making our approach a valuable near-to-intermediate term tool.

A potential bottleneck in the optimization of variational quantum algorithms are barren plateaus \cite{McClean18,larocca2024}, i.e., the exponential concentration of the loss function with increasing system size. However, we can readily incorporate strategies from existing research to mitigate their effects, should they start to appear at larger system sizes, within our framework. These include using shallow quantum circuits and local observables \cite{Cerezo21}, equivariant ansatze \cite{Schatzki24}, or smart initialization strategies \cite{puig2024,Park2024,Rudolph23}.

An important next step of this work would be to run the parameterized quantum circuits on actual quantum hardware, allowing us to access larger system sizes and investigate the scaling of the algorithm with the number of particles. Furthermore, we could analyze the effects of hardware noise which still largely limits the capabilities of current devices. However, given the ongoing advances in reducing hardware noise, we anticipate that our approach will soon become advantageous, enabling us to achieve system sizes of practical interest. 
Another interesting direction for future research would be to generalize the ideas presented in this work to the problem of real-time evolution \cite{Carleo12,Carleo17b}. This would allow simulating the non-equilibrium properties of continuous-space systems using quantum resources without the need for discretization.

\begin{acknowledgments}
We would like to thank Gian Gentinetta, Stefano Barison, Jannes Nys, Supanut Thanasilp, David Linteau, Markus Holzmann, Zakari Denis, and Clemens Giuliani for insightful discussions. 
This research was supported by the NCCR MARVEL, a National Centre of Competence in Research, funded by the Swiss National Science Foundation (grant number 205602).
\end{acknowledgments}

\clearpage

\appendix
\begin{widetext}

\section{Details on the algorithm \label{app:vmc}}

The pseudocode of the VMC optimization routine is shown in Algorithm~\ref{alg}, with $\Psi_\gamma(\mathbf{R}) = \exp(\braket{0|U_\gamma^\dagger(\mathbf{R})\hat{O}U_\gamma(\mathbf{R})|0})$ as an exemplary PQC ansatz wave function. All steps involving a quantum computation are highlighted in blue. Note that the use of quantum resources for the Monte Carlo sampling routine can be entirely removed by employing importance sampling [see Section~\ref{sec:is}]. In this case, the algorithm has to be adapted accordingly.

\begin{algorithm}[H]
\caption{VMC optimization with PQC ansatz (calculations on quantum devices are highlighted)}\label{alg}
\begin{algorithmic}
\Require Hamiltonian $H=T+V=-\frac{1}{2} \sum_{i=1}^N \nabla_{\mathbf{r}_i}^2+V(\mathbf{R})$
\Require Wave function ansatz, e.g., $\Psi_{\gamma}(\mathbf{R}) = \exp(\phi_{\gamma}(\mathbf{R})) = \exp(\braket{0|U_\gamma^\dagger(\mathbf{R})\hat{O}U_\gamma(\mathbf{R})|0})$
\For{step in $N_{\mathrm{iterations}}$}
\State  \Comment{Sample set of $M$ particle configurations $\{\mathbf{R}^{[1]},\dots, \mathbf{R}^{[M]}\}$ using MCMC}
\State List of samples $=\{\}$
\State Set initial configuration $\mathbf{R}$ (sampled randomly or final configuration from previous iteration)
\While{number of samples $<M$}
\State Generate candidate configuration $\mathbf{R}' = \mathbf{R} + \mathcal{N}(0,\sigma)$
\State \hl{Compute $\phi_{\gamma}(\mathbf{R}') = \braket{0|U_\gamma^\dagger(\mathbf{R}')\hat{O}U_\gamma(\mathbf{R}')|0}$ on quantum device}
\State Sample $u\sim \mathcal{U}[0,1]$
\If{$u<\min\left(1, \frac{|\exp(\phi_{\gamma}(\mathbf{R}'))|^2}{|\exp(\phi_{\gamma}(\mathbf{R}))|^2}\right)$}
\State Accept candidate configuration: $\mathbf{R}\gets \mathbf{R}'$
\State Append $\mathbf{R}$ to list of samples
\EndIf
\EndWhile
\State \Comment{Compute local energy $E_{\mathrm{loc}}(\mathbf{R})$ and gradients $O_\alpha(\mathbf{R})$ on each sample}
\For{$\mathbf{R}$ in list of samples}
\For{all parameters $\gamma_{\alpha}$}
\State \hl{Evaluate $O_\alpha(\mathbf{R})=\partial \gamma_\alpha \phi_{\gamma}(\mathbf{R}) = \partial \gamma_\alpha \braket{0|U_\gamma^\dagger(\mathbf{R})\hat{O}U_\gamma(\mathbf{R})|0}$ on quantum device using, e.g., parameter shift rule}
\EndFor
\State \hl{Evaluate $T_{\mathrm{loc}}(\mathbf{R})=-\frac 1 2 \sum_i \left(\nabla_{\mathbf{r}_i} \phi_{\gamma}(\mathbf{R})\right)^2 + \left(\nabla_{\mathbf{r}_i}^2\phi_{\gamma}(\mathbf{R})\right)$ on quantum device using, e.g., parameter shift rule}
\State Compute $E_{\mathrm{loc}}(\mathbf{R}) = T_{\mathrm{loc}}(\mathbf{R}) + V(\mathbf{R})$
\EndFor
\State \Comment{Compute force vector $F_{\alpha}$ and QGT $S_{\alpha\beta}$ via Monte-Carlo integration}
\For{all parameters $\gamma_{\alpha}, \gamma_{\beta}$}
\State Compute averages $\braket{E_{\mathrm{loc}} O_\alpha}, \braket{E_{\mathrm{loc}}}, \braket{O_\alpha}, \braket{O_\alpha^\dagger}, \langle O_\alpha^{\dagger} O_\beta\rangle$ over samples $\{\mathbf{R}^{[1]},\dots, \mathbf{R}^{[M]}\}$
\State Evaluate $F_\alpha = 2 \left(\braket{E_{\mathrm{loc}} O_\alpha} - \braket{E_{\mathrm{loc}}} \braket{O_\alpha}\right)$
\State Evaluate $S_{\alpha, \beta}=\left\langle O_\alpha^{\dagger} O_\beta\right\rangle-\left\langle O_\alpha^{\dagger}\right\rangle\left\langle O_\beta\right\rangle$
\EndFor
\State \Comment{Update model parameters $\gamma_\alpha$ via stochastic reconfiguration}
\State Invert QGT $S_{\alpha, \beta}$ using, e.g., SVD decomposition
\For{all parameters $\gamma_{\alpha}$}
\State Update parameter $\gamma_\alpha \gets \gamma_\alpha -\eta S^{-1}_{\alpha, \beta} F_{\beta}$ 
\EndFor
\EndFor
\end{algorithmic}
\end{algorithm}

The hyperparameters used for simulating the different systems of the main text are shown in Table~\ref{table:hyp}. The PQC circuit parameters are initialized as follows.

\textbf{Quantum rotor model:} The parameters in the encoding unitary are randomly sampled from a Gaussian distribution with mean 1 and standard deviation 0.01, i.e.~$\mathcal{N}(\mu=1, \sigma=0.01)$. The parameters in the purely variational 1-and 2-qubit unitaries are drawn from $\mathcal{N}(\mu=0, \sigma=1.0)$ and the classical parameters from $\mathcal{N}(\mu=1, \sigma=0.01)$.

\textbf{Helium-3 and homogeneous electron gas:}~The parameters in the encoding unitary are initialized to $0.5 \tilde{d}\left(x_i, x_j\right) + \gamma_b^{(l)}$ where $\gamma_b^{(l)}$ is drawn from $\mathcal{N}(\mu=0, \sigma=1.0)$. The parameters in the purely variational 1-and 2-qubit unitaries are sampled from $\mathcal{N}(\mu=0, \sigma=1.0)$ and the classical parameters from $\mathcal{N}(\mu=0, \sigma=0.1)$.

\begin{table}[h!]
\centering
\begin{tabular}{ |m{4cm}||m{3.0cm}|m{3.0cm}|m{3.0cm}|  }
 \multicolumn{4}{c}{(a) Quantum rotor model} \\
 \hline
 Hyperparameters & PQC$_1$ ($N=4$) & PQC$_2$ ($N\!=\!2,\!\dots\!,8$) & hybrid PQC$_2$ ($N\!=\!4$) \\
 \hline\hline
 $N_\text{iterations}$  & 500  & $500-1000$ &  1000 \\ \hline
 Number of MC samples $M $&   $10^4$  & $10^3 - 10^4$  & $10^4$ \\ \hline
 Learning rate $\eta$ &   0.05  & $0.05-0.1$  & 0.1 \\ \hline
 MCMC proposal dist. $\sigma$ &   0.5  & 0.5  & 0.5 \\ \hline
\end{tabular}
\centering
\begin{tabular}{ |m{4cm}||m{3.0cm}|m{3.0cm}|  }
  \multicolumn{3}{c}{\hspace*{3.5cm} (b) Helium-3}\\
 \hline
 Hyperparameters & hybrid PQC$_2$ $d=1$ & hybrid PQC$_2$ $d=2$ \\
 \hline\hline
 $N_\text{iterations}$  & 2000  & $4000$  \\ \hline
 Number of MC samples $M $&   $10^4$  & $4096$  \\ \hline
 Learning rate $\eta$ &   0.0005  & $0.001$  \\ \hline
 MCMC proposal dist. $\sigma$ &   0.1  & 0.2 \\ \hline
\end{tabular}
\quad
\begin{tabular}{ |m{3.0cm}|  }
  \multicolumn{1}{c}{(c) electron gas}\\
 \hline
 hybrid PQC$_2$ $d=2$ \\
 \hline\hline
  $2000$  \\ \hline
 $4096$  \\ \hline
 $0.5$  \\ \hline
  0.1 \\ \hline
\end{tabular}

\caption{Optimization hyperparameters used for training the PQC ansatze.}
\label{table:hyp}
\end{table}

\subsection{Computing gradients \label{app:param_shift}}
The VMC optimization in continuous space requires us to compute gradients of the log-amplitudes $\phi_\gamma(\mathbf{R}) = \log \Psi_\gamma (\mathbf{R})$ with respect to all parameters $\gamma_i$ and with respect to all particle positions $\mathbf{r}_i$. In either case, we can differentiate the log-amplitudes using the chain and product rules until we reach expressions that involve derivatives of the quantum computation output, that is, expectation values of observables $\langle \hat{O} \rangle_{\mathbf{R},\gamma} = \braket{0|U_\gamma^\dagger(\mathbf{R})\hat{O}U_\gamma(\mathbf{R})|0}$. If each parameter $\gamma_i$ enters as the angle of a single parameterized (multi-qubit) rotation gate, i.e., $\exp(-i\gamma_i \hat{P}/2)$, we can use the parameter shift rule \cite{Mitarai18,Schuld19} to obtain an expression for the gradient that is exact up to shot noise
\begin{align}
    \partial_{\gamma_i} \langle\hat{O} \rangle_{\mathbf{R},\gamma} = \frac 1 2 \left[ \langle\hat{O} \rangle_{\mathbf{R},\gamma;\gamma_i+\frac \pi 2} - \langle\hat{O} \rangle_{\mathbf{R},\gamma;\gamma_i-\frac \pi 2}\right].
\end{align}
Thus, a derivative amounts to calculating two distinct expectation values where the corresponding parameter of the gate is shifted by $+\frac \pi 2$ and $-\frac \pi 2$ respectively.

Similarly, if the particle coordinates $\mathbf{r}_i = (x_i,y_i,z_i)$ enter as the angle of a single parameterized gate through some function $\alpha(\mathbf{r}_i)$, i.e., $\exp(-i \alpha(\mathbf{r}_i) \hat{P}/2)$, the parameter shift rule can be used to compute each element of the gradient separately. For example,
\begin{align}
    \partial_{x_i} \langle\hat{O} \rangle_{\mathbf{R},\gamma} = \frac 1 2 \left[ \langle\hat{O} \rangle_{\mathbf{R},\gamma;\alpha(\mathbf{r}_i)+\frac \pi 2} - \langle\hat{O} \rangle_{\mathbf{R},\gamma;\alpha(\mathbf{r}_i)-\frac \pi 2}\right] \partial_{x_i} \alpha(\mathbf{r}_i).
\end{align}
When calculating the Laplacian the parameter shift rule can be applied twice to yield
\begin{align}
    \partial^2_{x_i} \langle\hat{O} \rangle_{\mathbf{R},\gamma} =  \frac 1 4 \left[ \langle\hat{O} \rangle_{\mathbf{R},\gamma;\alpha(\mathbf{r}_i)+\pi} + \langle\hat{O} \rangle_{\mathbf{R},\gamma;\alpha(\mathbf{r}_i)- \pi} - \langle\hat{O} \rangle_{\mathbf{R},\gamma}\right] \partial_{x_i} \alpha(\mathbf{r}_i) + \frac 1 2 \left[ \langle\hat{O} \rangle_{\mathbf{R},\gamma;\alpha(\mathbf{r}_i)+\frac \pi 2} - \langle\hat{O} \rangle_{\mathbf{R},\gamma;\alpha(\mathbf{r}_i)-\frac \pi 2}\right] \partial^2_{x_i} \alpha(\mathbf{r}_i) .
\end{align}
However, in many of the considered ansatz circuits, the parameters $\gamma_i$ (positions $\mathbf{r}_i$) enter as angles of multiple gates. For example, in the circuits used for simulating fermionic systems, several parameters within a layer are shared to fulfill the equivariance condition. Furthermore, the encoding of each particle position is typically repeated over several layers to increase the ansatz expressivity. This results in multiple unitaries depending on the same parameters (positions). When computing gradients, we can employ the product rule and for each occurrence of the parameter (position) in the circuit obtain a separate parameter shift rule. Following this simple recipe for the ansatz circuits considered in this work, we arrive at a required number of circuit evaluations as listed in Table~\ref{table:1}.

\begin{table}[h!]
\centering
\begin{tabular}{ |m{4cm}||m{3.5cm}|m{4.3cm}|m{4.3cm}|  }
 \multicolumn{4}{c}{(a) Gradients w.r.t. parameters $\gamma_i$} \\
 \hline
 type of gates & number of gates & distinct circuits for $\partial \log \Psi$ & distinct circuits for $\partial^2 \log \Psi$\\
 \hline\hline
 1-qubit gates (e.g.~$e^{-i\frac{\gamma}{2}\hat X}$)   & $2dNL$    & $4dNL\sim dNL$ &   not required \\ \hline
 2-qubit gates (e.g.~$e^{-i\frac{\gamma}{2}\hat X \hat X}$)&   $\frac{dN(dN-1)}{2}L$  & $dN(dN-1)L \sim d^2N^2L$  & not required\\ \hline
 2-qubit encoding gates \newline (e.g.~$e^{-i\frac{\gamma}{2}\alpha(\mathbf{r})\hat X \hat X}$) & $d\frac{N(N-1)}{2} L$ & $dN(N-1)L \sim d N^2 L$&  not required\\
 \hline
\end{tabular}

\begin{tabular}{ |m{4cm}||m{3.5cm}|m{4.3cm}|m{4.3cm}|  }
 \multicolumn{4}{c}{(b) Gradients w.r.t. all particle positions $\mathbf{r}_i$ / rotor angles $\theta_i$} \\
 \hline
 type of gates & number of gates & distinct circuits for $\partial \log \Psi$ & distinct circuits for $\partial^2 \log \Psi$\\
 \hline\hline
 1-qubit encoding gates \newline (e.g.~$e^{-i\frac{\theta}{2}\hat X}$)   & $dNL$    & $2dNL\sim dNL$ &   $2dNL^2 + 1\sim dNL^2$ \\ \hline
 2-qubit encoding gates \newline (e.g.~$e^{-i\frac{\gamma}{2}\alpha(\mathbf{r})\hat X \hat X}$)&   $d\frac{N(N-1)}{2} L$  & $dN(N-1)L \sim dN^2L$  & $d\frac{N(N-1)(2N-1)}{3}L^2 \!+\! 1\sim dN^3 L^2$\\ \hline
\end{tabular}
\caption{Resource requirements for computing gradients with respect to wave function parameters $\gamma$ (a) and with respect to particle positions $\mathbf{r}$ (or rotor angles $\theta$) (b). The former is used to optimize the wave function ansatz, while the latter together with the second order derivative (Laplacian) is required for evaluating the energy of the state. For each different type of gate (encoding and variational), we display the number of gates in the circuit and the number of distinct circuits that has to be executed to compute the gradient and the Laplacian (if required). The resources depend on the dimension $d$ of the system, the number of particles $N$, and the number of circuit layers $L$. Note that to estimate the overall number of circuit evaluations, the number of distinct circuits has to be multiplied by the number of measurement shots, the number of Monte Carlo samples, and the number of optimization iterations.}
\label{table:1}
\end{table}

The resources depend on the specific choice of circuit ansatz and therefore different scalings can be realized by adapting the circuits accordingly. For the more expressive PQC$_2$ circuits defined in this paper, the resource requirements are dominated by the computation of the Laplacian for each particle position which scales as $dN^3L^2$. However, one order of $N$ in the scaling can be omitted, if, instead of summing over all particles in the kinetic energy term, we evaluate the gradients only for a single particle. 
Note also that the number of circuit layers $L$ has an implicit dependence on the system size $N$ and the spatial dimension $d$ as we have empirically shown in Figs.~\ref{fig:rotor1},\ref{fig:helium}. However, the exact scaling is difficult to determine and highly system-specific.

Overall, the number of circuit evaluations for computing gradients scales linearly with the number of parameters. In contrast, computing gradients of classical functions via backpropagation requires only a single forward and backward pass, since intermediate steps of the computation can be stored and reused. Recent works \cite{bowles2023,Abbas23} have addressed this scaling discrepancy and proposed more efficient schemes, for example using multiple copies of the quantum state \cite{Abbas23}. However, these approaches come with their own drawbacks, and it is still unclear whether efficient training (in the classical sense) is possible. An entirely different approach to avoiding these optimization challenges presents the emerging fields of quantum extreme learning and quantum reservoir computing \cite{Fujii17,Mujal21,Weijie23,Innocenti23}. Here, the quantum computation does not depend on any trainable parameters but rather pre-processes the data for a subsequent classical machine learning pipeline. Such a scheme can be regarded as yet another example of hybrid quantum-classical ansatze which can be explored within our framework.

\subsection{Effects of shot noise \label{app:noise}}

The different wave function ansatze introduced in this work involve computing expectation values of observables by sampling from a quantum computer and thus, are necessarily subject to noise. The standard deviation $\sigma$ of an estimated expectation value of an observable $\hat{O}$ is given by
\begin{equation}
    \sigma_{\hat{O}}^2 = \frac{\mathrm{Var}[\hat{O}]}{N_{\mathrm{shots}}} = \frac{1}{N_{\mathrm{shots}}} (\langle \hat{O}^2\rangle_{\mathbf{R},\gamma} - \langle \hat{O}\rangle_{\mathbf{R},\gamma}^2),
\end{equation}
where $\langle \hat{O} \rangle_{\mathbf{R},\gamma}$ is a shorthand for $\braket{0|U_\gamma^\dagger(\mathbf{R})\hat{O}U_\gamma(\mathbf{R})|0}$. These errors propagate through the computation of the wave function giving rise to a noisy estimate of the exact wave function amplitudes. Defining the ansatz as $\Psi(\mathbf{R}) = \exp(\phi_\gamma(\mathbf{R})) = \exp(\langle \hat{O}\rangle_{\mathbf{R},\gamma})$, the error on the wave function amplitudes is $\sigma_{\Psi} = |\Psi(\mathbf{R})| \sigma_{\hat{O}}$. However, often the relevant quantities of interest are not the wave function amplitudes themselves, but expectation values of observables such as the energy $E=\frac{\langle \Psi_\gamma| H|\Psi_\gamma \rangle}{\braket{\Psi_\gamma|\Psi_\gamma}}$. The error on the energy can be expressed as
\begin{equation}\label{eq:sigmaE}
    \sigma^2_E = \frac N M \left[ \frac 1 4 \sigma^2_{\partial^2} + \frac 1 N \sum_{i=1}^N \langle (\nabla_{\mathbf{r}_i} \phi_\gamma(\mathbf{R}))^2 \rangle \sigma_\partial^2 \right],
\end{equation}
where $N$ is the number of particles and $M$ the number of Monte Carlo samples. Here, $\langle \cdot \rangle$ denotes the average taken over all Monte Carlo sampled positions. Finally, $\sigma^2_{\partial}$ and $\sigma^2_{\partial^2}$ are the variances of the log-amplitude gradients and Laplacian respectively. If all gradients are computed via the parameter shift rule, resulting in a sum of expectation values, the variances $\sigma^2_{\partial}$ and $\sigma^2_{\partial^2}$ will roughly scale with the number of distinct expectation values (i.e.~circuits) [see Table~\ref{table:1}]. This scaling argument, however, assumes that the observables are uncorrelated and give rise to a similar variance for each circuit which, in general, will not be the case. Nevertheless, the number of distinct circuits to execute provides a rough estimate for how errors accumulate. Finally, note that the variance of the energy in Eq.~\eqref{eq:sigmaE} is bound as long as the gradient of the log-amplitudes is bound at all sampled particle positions which is a reasonable assumption.

In the case of fermionic wave functions we proposed an ansatz given by a Slater determinant of single-particle plane wave orbitals with backflow transformations [see Section~\ref{sec:backflow}] of the form
\begin{equation}
    q_{ir,\gamma}(\mathbf{R}) =  x_{ir} + \langle \hat{O}_{ir} \rangle_{\mathbf{R}, \gamma_\mathrm{real}} + i  \langle \hat{O}_{ir} \rangle_{\mathbf{R}, \gamma_\mathrm{imag}},
\end{equation}
where $i$ indexes the particle, and $r$ the spatial dimension. For simplicity we again assume the variances $\sigma_{\hat O}^2$ of all observables to be roughly the same. In this case the variance of the real and complex elements of the wave function amplitudes can be estimated as
\begin{equation}
    \sigma^2_\Psi = N \sigma_{\hat O}^2 \left[\frac 1 N \sum_{j=1}^N ||\mathbf{k}_j||^2\right] \frac{1}{N!} \sum_{\sigma \in S_N} \prod_{i=1}^N e^{-\mathbf{k}_{\sigma(i)}\cdot (\mathbf{q}_i - \mathbf{q}^*_i)},
\end{equation}
where, for brevity and clarity of notation, we have further assumed that all covariances appearing in the error propagation vanish.

\section{Quantum rotor model \label{app:rotor}}
\subsection{Discretized model \& MPS benchmarks  \label{app:discrete}}
The quantum rotor Hamiltonian [Eq.~\eqref{eq:rotor}] can be expressed in a discrete basis spanned by the eigenvectors $\{\ket{m}\}$ of the angular momentum operator $L_i=-i \partial_{\theta_i}$, i.e., $\langle\theta_i | m_i\rangle=\frac{e^{-i m_i \theta_i}}{\sqrt{2 \pi}} $ with $ m_i \in \mathbb{Z} $ and $ i\in\{1,\dots, N\}$. For the many-body state of $N$ rotors, we can choose a basis of product states $|\mathbf{m}\rangle=\left|m_1, \ldots, m_N\right\rangle$. In this basis Eq.~\eqref{eq:rotor} takes the form
\begin{equation}
    H=\frac{1}{2} \sum_{i=1}^N L_i^2-\frac{1}{2} \sum_{i=1}^{N-1}\left(L_i^{+} L_{i+1}^{-}+L_{i+1}^{+} L_i^{-}\right),
\end{equation}
where
\begin{equation}
    L_i^{+} \equiv \sum_{m_i \in \mathbb{Z}}\left|m_i+1\right\rangle\left\langle m_i\right| \quad \text { and } \quad L_i^{-} \equiv \sum_{m_i \in \mathbb{Z}}\left|m_i-1\right\rangle\left\langle m_i\right| ,
\end{equation}
with $\left[L_i^{+}, L_i^{-}\right]=0$.

To numerically treat the rotor model, we truncate the infinitely-sized local basis of each rotor to some maximum angular momentum cutoff $M$: $\{|-M\rangle, \ldots,|M\rangle\}$. The resultant many-body Hilbert space has a finite dimension of $(2M+1)^N$.

\begin{figure*}[t!]
    \centering
    \includegraphics[width=0.9\textwidth]{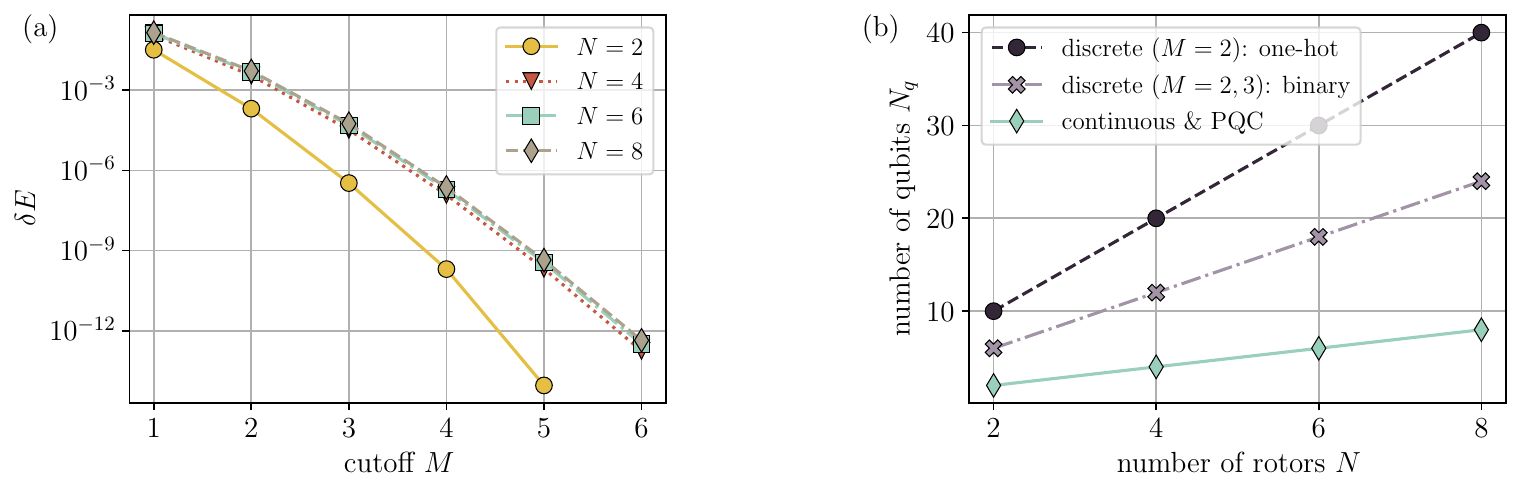}
    \caption{Quantum rotor model --- {\bf (a)} MPS simulations of the discretized rotor model for different angular momentum cutoffs $M$ of the local Hilbert spaces. Shown are the relative errors to the exact ground state energy for different system sizes $N$. {\bf (b)} Number of qubits required to simulate a $N$-rotor system on a quantum device using an approach based on discretization (circle, crosses) and the continuous-space framework introduced in this work (diamonds).
    }
    \label{fig:app:rotor}
\end{figure*}

In the following, we use this discretized model for two purposes: First, we obtain the exact ground state energy of the rotor system to benchmark the achieved accuracies of the VMC results reported in Section~\ref{sec:rotor} of the main text. Second, we determine the minimum number of qubits that would be required by a quantum simulation of the rotor model to achieve a certain accuracy. These resources are then compared to those of our continuous-space framework.

To determine the exact ground-state energy of the system we employ matrix product states (MPS)~\cite{white92,schollwock11,orus14} together with the density-matrix renormalization group algorithm (DMRG) provided by the ITensor library \cite{itensor}. We perform simulations for different system sizes and a range of angular momentum cutoffs $M$, while keeping the MPS bond dimension large enough to consider all results exact up to basis truncation errors. If $E(M=k)-E(M=k+1) \leq 10^{-14}$ we define the exact ground state energy as $E_{\mathrm{exact}} = E(M=k)$. In Fig.~\ref{fig:app:rotor}(a) we plot the relative error w.r.t. the exact energy, $\delta E = |E(M)-E_{\mathrm{exact}}|/|E_{\mathrm{exact}}|$ as a function of the momentum cutoff $M$.

On a digital quantum device, we need to map the states and operators of the $(2M+1)^N$ dimensional rotor Hilbert space to a qubit representation with dimension $2^{N_q}$, where $N_q$ is the number of qubits on the quantum device. There are two common choices of encoding a $(2M+1)$-level system into a two-level system: the one-hot encoding, which uses $(2M+1)$ qubits per rotor and the binary or gray encodings, requiring only $\lceil \log_2(2M+1) \rceil$ qubits per rotor. Importantly, the total number of qubits $N_q$ depends linearly on the number of rotors $N$ and linearly (logarithmically) on the angular momentum cutoff $M$. The accuracy as a function of the cutoff $M$ reported in Fig.~\ref{fig:app:rotor}(a) provides an upper bound for the achievable accuracy on the quantum computer. To determine the necessary number of qubits on the quantum device $N_q$, we compare the MPS accuracies with the accuracy obtained using the VMC framework, which provides an estimate for the required cutoff $M$ and consequently for $N_q$. 

From Fig.~\ref{fig:rotor1}(c) of the main text we observe that a 2-layer PQC surpasses the energy of a $M=2$ MPS simulation, while 10 layers are required to reach the accuracy of a $M=3$ simulation. In Fig.~\ref{fig:app:rotor}(b) we show the estimated number of qubits $N_q$ to simulate the system for different sizes $N$ for $M=2,3$. 
While the scaling of all considered quantum approaches is linear in system size, our continuous-space method profits from a drastically reduced prefactor and, consequently, a reduction in the number of qubits, which is particularly beneficial for near-term quantum devices. For example, our scheme allows us to simulate a $N=8$ rotor system with 8 qubits instead of 24, that would be required to achieve the same accuracy conventionally.

Note that the accuracy for a chosen basis and Hilbert space cutoff $M$ depends on the model parameters, i.e., the relative strength between kinetic and potential energy, which we have set to unity. If the kinetic energy dominates, the plane wave basis with a small cutoff will provide highly accurate results, while in the opposite regime a larger number of angular momentum eigenstates have to be taken into account to match the same accuracy. This directly affects the required number of qubits. In general, choosing an optimal basis that allows an efficient use of quantum resources is non-trivial. Moreover, simulations have to be performed at different Hilbert space cutoffs to estimate errors and extrapolate to the continuum limit. Our framework, on the other hand, does not rely on choosing a specific discrete basis and the number of qubits is always fixed irrespective of the model parameters.

\subsection{Comparison of different encoding and variational ansatze \label{app:comparison}}

\begin{figure*}[t!]
    \centering
    \includegraphics[width=0.99\textwidth]{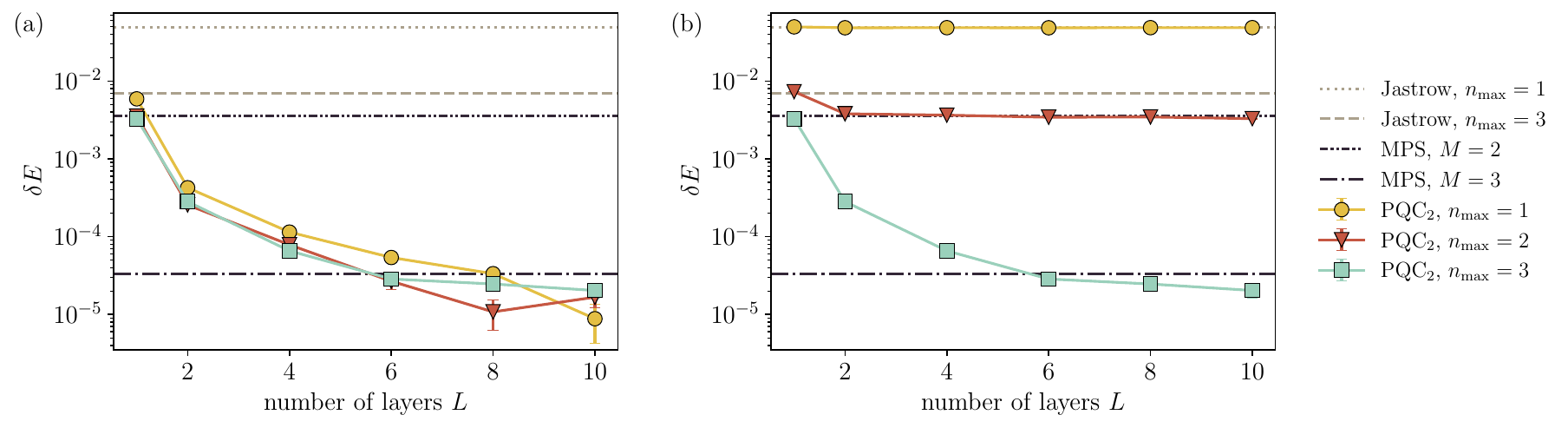}
    \caption{Quantum rotor model --- Comparing the expressive power of different coupling maps underlying the variational unitaries {\bf (a)} and encoding unitaries {\bf (b)}. Shown are the relative errors to the exact ground state energy obtained on the $N=4$ rotor system. $n_{\text{max}} =1$ corresponds to a nearest-neighbor coupling (yellow circles), for $n_{\text{max}} =2$ additional next nearest-neighbor gates are added (red triangles), and $n_{\text{max}} =3$ corresponds to an all-to-all entangling map (blue squares). {\bf (a)} For the variational unitary, parameterized nearest-neighbor gates suffice to reach high accuracies which can be controlled by the number of circuit layers $L$. {\bf (b)} In contrast, we require an all-to-all coupling map for the encoding unitary in order to obtain an expressive and controllable ansatz.
    }
    \label{fig:app:rotor2}
\end{figure*}

In the following we investigate the effects of different encoding and variational circuit structures on the performance of the PQC ansatz. As a starting point we consider the PQC$_2$ circuit introduced in Section~\ref{sec:rotor} with the two-body rotor features $\cos(\theta_i-\theta_{i+n})$ encoded into the angles of two-qubit rotation gates acting on the respective qubits $i$ and $i+n$. Similarly, the variational unitary is composed of parameterized two-qubit rotation gates, however, in this case they are acting on neighboring qubits only. We now examine whether the accuracy on the ground state energy can be improved by introducing additional parameterized long-range gates that connect distant qubits $i$ and $i+n$ with $n>1$. Taking the $N=4$ rotor system as an example, Fig.~\ref{fig:app:rotor2}(a) shows the achieved relative error to the exact ground state energy for three different ansatze with varying connectivities:~$n_{\text{max}}=1$ corresponds to the circuit of the main text involving only nearest-neighbor variational gates, for $n_{\text{max}}=2$ next-nearest neighbor gates are added, and $n_{\text{max}}=3$ represents an all-to-all entangling map for the $N=4$ rotor example. All circuits give rise to similar errors and roughly the same scaling with the number of layers $L$, suggesting that a hardware-efficient nearest-neighbor coupling map suffices for the variational unitary in this case.

Next, we fix the entangling map of the variational unitary to $n_{\text{max}}=3$ and instead vary the connectivity of the encoding gates. Thus, $n_{\text{max}}=1$ corresponds to the case where only nearest-neighbor rotor features are encoded, while for $n_{\text{max}}=3$ the two-body features between all pairs of rotors are embedded onto the quantum state. Fig.~\ref{fig:app:rotor2}(b) shows the achieved accuracies for the three optimized circuits. The PQCs with $n_{\text{max}}=1,2$ perform considerably worse than the $n_{\text{max}}=3$ PQC. Furthermore, the $n_{\text{max}}=1,2$ results for the energy do not improve by increasing the number of layers beyond $L=2$. Interestingly, a PQC with $n_{\text{max}}=1$ has the same expressive power as a classical Jastrow ansatz with nearest-neighbor interactions, while the achieved energy of the PQC with $n_{\text{max}}=2$ converges close to the energy obtained from a MPS simulation with a local Hilbert space cutoff $M=2$. Overall, we conclude that in contrast to the variational unitary, we require an all-to-all connection for the encoding unitary in order to have a controllable and expressive ansatz. Note however, that this observation might change depending on the physical system to be studied, the nature of the interactions, and the chosen feature map of the encoding.

\section{Helium-3 \label{app:helium}}
\subsection{Aziz potential \label{app:aziz}}
The Aziz potential modeling the inter-particle interactions of $^3$He is given by \cite{aziz77}
\begin{equation}
    V(x) = \epsilon \left[A \exp(-\alpha x) - \left(\frac{C_6}{x^6} + \frac{C_8}{x^8} + \frac{C_{10}}{x^{10}}\right) F(x) \right] ,
\end{equation}
with
\begin{equation}
    F(x)=
\begin{cases}
\exp\left(-\left(\frac{D}{x}-1\right)^2\right) \quad & \text{for } x < D\\
1  & \text{for } x\geq D
\end{cases} .
\end{equation}
The parameters are shown in Table~\ref{table:aziz}.

\begin{table}[h!]
\centering
\begin{tabular}{ |c|c|c|c|c|c|c|  }
\hline
 $\epsilon$ & $A$ & $\alpha$ & $C_6$ & $C_8$ & $C_{10}$ & $D$\\
 \hline
 7.846373   & $0.544850 \times 10^6$    & $13.353384$ &   1.37332412 & 0.4253785 & 0.178100 & 1.241314\\
 \hline
\end{tabular}
\caption{Parameters of the Aziz potential.}
\label{table:aziz}
\end{table}

\subsection{Equivariant circuits \label{app:equivariant}}

In the following we show that the PQCs used in Section~\ref{sec:helium} are equivariant with respect to particle exchanges, thus ensuring the antisymmetry of the fermionic wave function ansatz.

The equivariance condition for the backflow transformation $\mathbf{R} \rightarrow \mathbf{Q}(\mathbf{R})$ can be expressed as $T_{ij} \mathbf{Q}(\mathbf{R}) = \mathbf{Q}(T_{ij}\mathbf{R}), \, \forall i\neq j$ where $T_{ij}$ denotes the transposition of particles $i,j$, i.e.
\begin{equation}
    T_{ij}(\mathbf{r}_1,\dots,\mathbf{r}_i,\dots, \mathbf{r}_j, \dots, \mathbf{r}_N) = (\mathbf{r}_1,\dots,\mathbf{r}_j,\dots, \mathbf{r}_i, \dots, \mathbf{r}_N).
\end{equation}
In terms of the individual quasi-particle positions $\mathbf{Q} = (\mathbf{q}_1,\dots, \mathbf{q}_N)$, equivariance implies
\begin{equation}
    \mathbf{q}_j(\mathbf{R}) = \mathbf{q}_i(T_{ij}\mathbf{R}) \quad \text{and} \quad \mathbf{q}_k(\mathbf{R}) = \mathbf{q}_k(T_{ij}\mathbf{R}), \quad \forall i\neq j\neq k.
\end{equation}
In Section~\ref{sec:helium} the backflow transformation $\mathbf{q}_i(\mathbf{R})$ is represented via a set of expectation values evaluated on a parameterized quantum circuit $\braket{0|U^\dagger_\gamma(\mathbf{R}) \hat{\mathbf{O}}_i U_\gamma(\mathbf{R})|0}$. We thus require
\begin{align}
    \braket{0|U^\dagger_\gamma(\mathbf{R}) \hat{\mathbf{O}}_j U_\gamma(\mathbf{R})|0} &= \braket{0|U^\dagger_\gamma(T_{ij}\mathbf{R}) \hat{\mathbf{O}}_i U_\gamma(T_{ij}\mathbf{R})|0} , \quad \forall i\neq j, \label{eq:app:eq1}\\
    \braket{0|U^\dagger_\gamma(\mathbf{R}) \hat{\mathbf{O}}_k U_\gamma(\mathbf{R})|0} &= \braket{0|U^\dagger_\gamma(T_{ij}\mathbf{R}) \hat{\mathbf{O}}_k U_\gamma(T_{ij}\mathbf{R})|0} , \quad \forall i\neq j\neq k. \label{eq:app:eq2}
\end{align}
The global unitary can be decomposed into variational and encoding unitaries as $U_\gamma(\mathbf{R}) = \prod_{l=1}^L V^{[l]}_\gamma E^{[l]}_\gamma(\mathbf{R})$. The parameters $\gamma$ appearing in the encoding unitary do not depend on the particle index and thus are shared among different encoding gates in a given layer $l$. Therefore, it follows that $E^{[l]}_\gamma(T_{ij}\mathbf{R}) = \text{SWAP}_{ij} E^{[l]}_\gamma(\mathbf{R}) \text{SWAP}_{ij}$ where $\text{SWAP}_{ij}$ is the SWAP gate exchanging qubits associated to the particle indices $i$ and $j$. Similarly, the parameters of the variational unitary are independent of the particle index and thus, the variational unitary is invariant to any particle transpositions $V^{[l]}_\gamma = \text{SWAP}_{ij} V^{[l]}_\gamma \text{SWAP}_{ij}$. Using $(\text{SWAP}_{ij})^2 = \mathbb{I}$, we can therefore write
\begin{align}
    U_\gamma(T_{ij}\mathbf{R})\ket{0} &= \prod_{l=1}^L V^{[l]}_\gamma E^{[l]}_\gamma(T_{ij}\mathbf{R})\ket{0}\\
    &= \prod_{l=1}^L V^{[l]}_\gamma \text{SWAP}_{ij} E^{[l]}_\gamma(\mathbf{R}) \text{SWAP}_{ij}\ket{0}\\
    & = \text{SWAP}_{ij}\left[\prod_{l=1}^L V^{[l]}_\gamma E^{[l]}_\gamma(\mathbf{R})\right]\text{SWAP}_{ij} \ket{0}\\
    & = \text{SWAP}_{ij} U_\gamma(\mathbf{R})\ket{0}
\end{align}
Hence, for the expectation values we find 
\begin{align}
    \braket{0|U^\dagger_\gamma(T_{ij}\mathbf{R}) \hat{\mathbf{O}}_i U_\gamma(T_{ij}\mathbf{R})|0} &= \braket{0|U^\dagger_\gamma(\mathbf{R}) \text{SWAP}_{ij}\hat{\mathbf{O}}_i\text{SWAP}_{ij} U_\gamma(\mathbf{R})|0} = \braket{0|U^\dagger_\gamma(\mathbf{R}) \hat{\mathbf{O}}_j U_\gamma(\mathbf{R})|0}, \quad \forall i\neq j, \\
    \braket{0|U^\dagger_\gamma(T_{ij}\mathbf{R}) \hat{\mathbf{O}}_k U_\gamma(T_{ij}\mathbf{R})|0} &= \braket{0|U^\dagger_\gamma(\mathbf{R}) \text{SWAP}_{ij}\hat{\mathbf{O}}_k\text{SWAP}_{ij} U_\gamma(\mathbf{R})|0} = \braket{0|U^\dagger_\gamma(\mathbf{R}) \hat{\mathbf{O}}_k U_\gamma(\mathbf{R})|0}, \quad \forall i\neq j\neq k,
\end{align}
which coincides with the equivariance conditions of Eqs.~\eqref{eq:app:eq1}-\eqref{eq:app:eq2}. Therefore, we have shown that our PQC ansatze give rise to equivariant backflow transformations.

\section{Homogeneous electron gas \label{app:heg}}

Another important fermionic system is the homogeneous electron gas (HEG) which is a paradigmatic model for studying the electronic properties of solids, in particular Alkali metals \cite{pines1963}. The HEG accounts for the Coulomb repulsion between the electrons in the solid, however, the interactions with the positively charged ions are modeled via a uniform, static, positive background. The two-dimensional HEG Hamiltonian can be expressed (in units of Hartree) as \cite{Tanatar89}
\begin{equation}\label{eq:heg}
    H=-\frac{1}{2 r_s^2} \sum_{i=1}^N \nabla_{\mathbf{r}_i}^2+\frac{1}{r_s} \sum_{i<j}^N \frac{1}{\left\|\mathbf{r}_i-\mathbf{r}_j\right\|}+\text { const. },
\end{equation}
where $r_s = \sqrt{1/(\pi n)}$ is the Wigner-Seitz radius, $n=N/L^2$ is the electron density, and $L$ is the size of the two-dimensional simulation cell subject to periodic boundary conditions. The constant term in Eq.~\eqref{eq:heg} arises from the electron-background interactions. The sum over the pairwise Coulomb interactions between electrons (and their periodic images) is carried out using the Ewald method which rewrites the original, conditionally convergent sum in terms of two absolutely convergent sums, one performed in real space and the other one in momentum space \cite{ewald,Abdulnour96,harris98,Wood04}.

In the following we consider a spin-polarized system of $N=3$ electrons. As in the case for Helium-3, we choose as the classical mean-field ansatz a Slater determinant $\Phi_D(\mathbf{R})$ of single-particle plane waves $\phi_\mu\left(\mathbf{r}_i\right) = \exp(i \mathbf{k}_\mu \cdot \mathbf{r}_i)$ with $\mathbf{k}_\mu=\frac{2 \pi}{L} \mathbf{n} $ and $\mathbf{n} \in \mathbb{Z}^d$. The mean-field wave function describes the system well only in the limit of small Wigner-Seitz radii $r_s \rightarrow 0$ where the kinetic energy dominates. In the regime of large interactions ($r_s\gg 1$), correlations between electrons become increasingly important. Therefore, we augment the mean-field state by a classical Jastrow factor $\Psi_{\text{S-J}}(\mathbf{R}) = J(\mathbf{R}) \Phi_D(\mathbf{R})$. We choose a form for the Jastrow that is commonly employed in Quantum Monte Carlo simulations of the HEG \cite{Whitehead16,Gabriel24}
\begin{equation}
    J(\mathbf{R})=\exp\left[\sum_{i<j} \sum_{n=1}^{N_j} c_{n}\left[j(x_{ij})^2+j(y_{ij})^2\right]^{n / 2}\right],
\end{equation}
where $x_{ij} = x_i - x_j$, $y_{ij} = y_i - y_j$ are computed via the minimum image convention and $j(x)=|x|\left[1-2(|x| / L)^3\right]$. The coefficients $c_n$ are trainable parameters and we fix the number of terms to $N_j=7$.

After pre-optimizing the classical Slater-Jastrow ansatz [see Fig.~\ref{fig:app:heg}(a)], we add quantum backflow transformations $\Psi_{\text{BF}}(\mathbf{R}) = J(\mathbf{R}) \Phi_D(\mathbf{Q}(\mathbf{R}))$ that are parameterized by a PQC [see Sec.~\ref{sec:helium}]. We use the the same ansatz as for the $d=2$ Helium system comprised of 6 qubits [c.f.~Fig.~\ref{fig:helium_circ}] and again encode the periodic surrogate distance between electrons into the quantum state. In Fig.~\ref{fig:app:heg}(b) we show a representative optimization curve for the electron gas at $r_s=200$ using a PQC backflow transformation with $L=4$ layers. The PQC backflow ansatz quickly improves on the pre-optimized Slater-Jastrow wave function (gray dashed line) reaching a final lower energy.

\begin{figure*}[t!]
    \centering
    \includegraphics[width=0.9\textwidth]{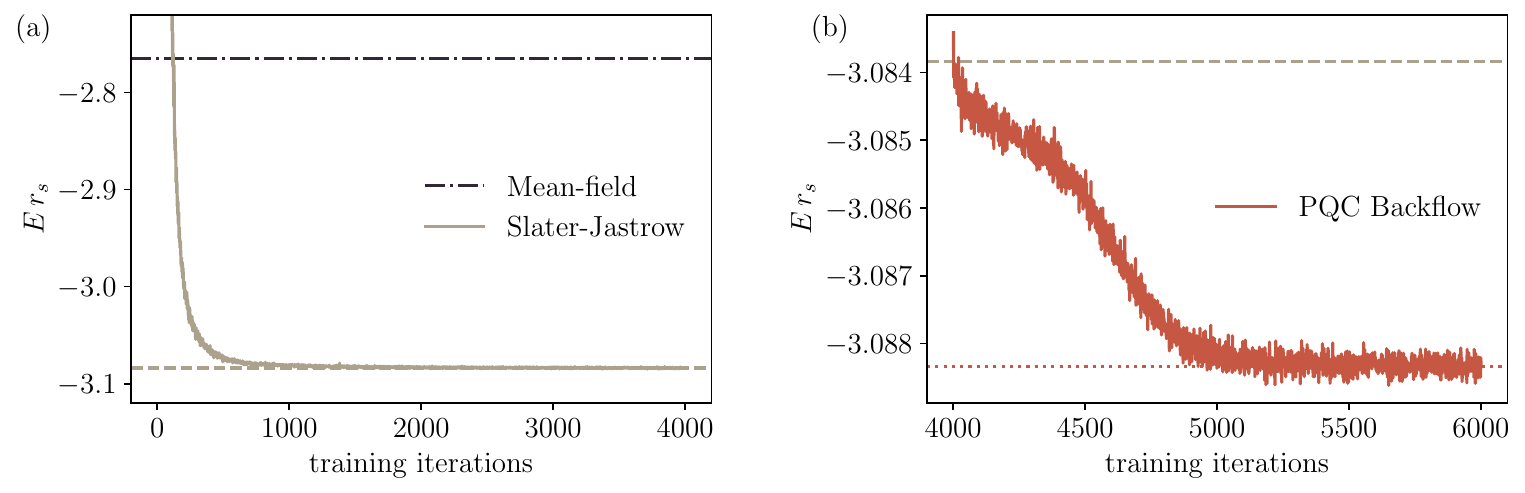}
    \caption{Homogeneous electron gas --- VMC optimization curves for a system of $N=3$ electrons in $d=2$ dimension with $r_s=200$. {\bf (a)} Energies (rescaled by $r_s$) of the classical Slater-Jastrow ansatz during training. The black dashed line denotes the energy of the Mean-field (Hartree-Fock) state. {\bf (b)} A PQC backflow transformation with $L=4$ layers is added to the pre-optimized Slater-Jastrow ansatz and training is continued such that the energy decreases further.
    }
    \label{fig:app:heg}
\end{figure*}

\end{widetext}

\bibliography{bibliography.bib}

\end{document}